\documentclass{aa}
\usepackage{natbib,psfig,graphicx}
\def\mathnew{\mathsurround=0pt}
\def\simov#1#2{\lower .5pt\vbox{\baselineskip0pt \lineskip-.5pt
\ialign{$\mathnew#1\hfil##\hfil$\crcr#2\crcr\sim\crcr}}}

\def\MeV{Me\kern-0.11em V}
\def\keV{ke\kern-0.11em V}

\begin{document}

\title{Intracluster light in  clusters of galaxies at redshifts $0.4<$z$<0.8$~\thanks{Based on observations made at ESO Telescopes at the Paranal 
Observatory under programme ID 082.A-0374. 
Also based on the use of the NASA/IPAC
Extragalactic Database (NED) which is operated by the Jet Propulsion
Laboratory, California Institute of Technology, under contract with the
National Aeronautics and Space Administration. Based on observations made with 
the NASA/ESA \textit{Hubble Space Telescope}, obtained from the data archives 
at the Space Telescope European Coordinating Facility and the Space Telescope 
Science Institute, which is operated by the Association of Universities for 
Research in Astronomy, Inc., under NASA contract NAS 5-26555.}}

\author{
L.~Guennou\inst{1} \and
C.~Adami\inst{1} \and
C.~Da Rocha\inst{2} \and
F.~Durret\inst{3,4} \and
M.P.~Ulmer\inst{5} \and
S.~Allam\inst{6} \and
S.~Basa\inst{1} \and
C.~Benoist\inst{7} \and
A.~Biviano\inst{8} \and
D.~Clowe\inst{9,10} \and
R.~Gavazzi\inst{3,4} \and
C.~Halliday\inst{11} \and
O.~Ilbert\inst{1} \and
D.~Johnston\inst{6} \and
D.~Just\inst{12} \and
R.~Kron\inst{13} \and
J.M.~Kubo\inst{6} \and
V.~Le Brun\inst{1} \and
P.~Marshall\inst{14,15} \and
A.~Mazure\inst{1} \and
K.J.~Murphy\inst{9} \and
D.N.E.~Pereira \inst{16} \and
C.R.~Raba\c ca \inst{16} \and
F.~Rostagni\inst{7} \and
G.~Rudnick\inst{17} \and
D.~Russeil\inst{1} \and
T.~Schrabback\inst{14,18} \and 
E.~Slezak\inst{7} \and
D.~Tucker\inst{6} \and 
D.~Zaritsky\inst{12}
}

\offprints{L.~Guennou \email{loic.guennou@oamp.fr}}

\institute{
LAM, OAMP, P\^ole de l'Etoile Site de Ch\^ateau-Gombert,
38 rue Fr\'ed\'eric Joliot-Curie, 13388 Marseille Cedex 13, France 
\and
N\'ucleo de Astrof\'{\i}sica Te\'orica, Universidade Cruzeiro do Sul, R. Galv\~ao Bueno 868, 01506--000, S\~ao Paulo, SP, Brazil  
\and
UPMC Universit\'e Paris 06, UMR~7095, Institut d'Astrophysique de Paris, 
F-75014, Paris, France 
\and
CNRS, UMR~7095, Institut d'Astrophysique de Paris, F-75014, Paris, France 
\and
Department Physics $\&$ Astronomy, Northwestern University, Evanston, 
IL 60208-2900, USA 
\and
Fermi National Accelerator Laboratory, P.O. Box 500, Batavia, IL 60510, USA 
\and
OCA, Cassiop\'ee, Boulevard de l'Observatoire, BP 4229, 06304 Nice Cedex 4, France 
\and
INAF/Osservatorio Astronomico di Trieste, via G. B. Tiepolo 11, I-34143, 
Trieste, Italy 
\and
Department of Physics and Astronomy, Ohio University, 251B Clippinger
Lab, Athens, OH 45701, USA 
\and
Alfred P. Sloan Fellow 
\and
Osservatorio Astrofisico di Arcetri, Largo Enrico Fermi 5, 50125 Firenze, Italy, 
\and
Steward Observatory, University of Arizona, 933 N. Cherry Ave. Tucson, 
AZ 85721, USA 
\and
Department of Astronomy and Astrophysics, The University of Chicago, 5640 South Ellis Avenue, Chicago, IL 60615, USA 
\and
Leiden Observatory, Leiden University, Niels Bohrweg 2, NL-2333 CA
Leiden, The Netherlands 
\and
Kavli Institute for Particle Astrophysics and Cosmology, Stanford University, 
2575 Sand Hill Road, Menlo Park, CA 94025, USA 
\and
Observatorio do Valongo, UFRJ, Ladeira do Pedro Antonio 43, Saude, Rio de Janeiro 20080-090, Rio de Janeiro, Brazil 
\and
Dept. of Physics and Astronomy, University of Kansas, Lawrence, KS 66045, USA 
\and
Kavli Institute for Particle Astrophysics and Cosmology, Stanford University, 382 Via Pueblo Mall, Stanford, CA 94305-4060, USA 
}

\date{Accepted . Received ; Draft printed: \today}

\authorrunning{Guennou et al.}

\titlerunning{Intracluster light in 0.4$<$z$<$0.8 clusters}

\abstract 
{The study of intracluster light (ICL) can help us to understand the
  mechanisms taking place in galaxy clusters, and to place constraints
  on the cluster formation history and physical properties. However,
  owing to the intrinsic faintness of ICL emission, most searches and
  detailed studies of ICL have been limited to redshifts z$<0.4$.}
{To help us extend our knowledge of ICL properties to
  higher redshifts and study the evolution of ICL with redshift, we
  search for ICL in a subsample of ten clusters detected by the
  ESO Distant Cluster Survey (EDisCS), at redshifts
  $0.4<$z$<0.8$, that are also part of our DAFT/FADA Survey.}
{We analyze the ICL by applying the OV WAV package, a wavelet-based
  technique, to deep HST ACS images in the F814W filter and to V-band VLT/FORS2
  images of three clusters. Detection levels are assessed as a function of the diffuse
  light source surface brightness using simulations. }
{ In the F814W filter images, we detect diffuse light sources in
  all the clusters, with typical sizes of a few tens of kpc (assuming
  that they are at the cluster redshifts). The ICL detected by
  stacking the ten F814W images shows an 8$\sigma$ detection
  in the source center extending over a $\sim 50\times 50$~kpc$^2$
  area, with a total absolute magnitude of $-21.6$ in the F814W
  filter, equivalent to about two $L^*$ galaxies per cluster. We find a weak
  correlation between the total F814W absolute magnitude of the ICL
  and the cluster velocity dispersion and mass. There is no apparent correlation 
  between the cluster mass-to-light ratio (M/L) and the amount of ICL, and no evidence for
  any preferential orientation in the ICL source distribution. We find no strong
  variation in the amount of ICL between z=0 and z=0.8. In addition, we
  find wavelet-detected compact objects (WDCOs) in the three clusters for which
  data in two bands are available; these objects are probably very faint compact
  galaxies that in some cases are members of the respective clusters and comparable
  to the faint dwarf galaxies of the Local Group.  }
{We have shown that ICL is important in clusters at least up to
  redshift z=0.8. The next step is now to detect it at even larger
  redshifts, to see if there is a privileged stage of cluster evolution where 
it has been
  stripped from galaxies and spread in the intracluster medium.}


\maketitle

\section{Introduction}

The search for intracluster light (hereafter ICL) provides a complementary
way of determining the mechanisms occurring inside galaxy clusters, as
well as constraining the properties and formation history of the ICL. These studies
promise to yield possible answers to many fundamental questions about the formation and
evolution of galaxy clusters and their constituent galaxies. In addition, it is important
to determine how and when the ICL formed, and the connection between the ICL and
the central brightest cluster galaxy (see e.g. Gonz\'alez et al., 2005).
Cosmological N-body and hydrodynamical simulations are beginning to predict the
kinematics and origin of the ICL (see e.g. Dolag et al. 2010). The ICL traces
the evolution of baryonic substructures in dense environments and can thus be used to
constrain some aspects of cosmological simulations that are uncertain, such as the
modeling of star formation and the mass distribution of the baryonic light-emitting
component in galaxies. The study by Da Rocha et al. (2005) also produced important
results about the significant presence of ICL in groups, which are crucial if
we assume that groups are the basic building blocks of clusters, that are able
to bring their own ICL to the cluster-building process.

From a technical point-of-view, modern CCD cameras now allow us to study the properties
of the  diffuse light in clusters, i.e. its morphology, radial distribution, and colors, in
a  quantitative way (e.g. Uson et al. 1991, Bernstein et al. 1995, Gregg \&
West 1998, Mihos et al. 2005, Zibetti et al. 2005, Gonz\'alez et al. 2007, Krick \&
Bernstein 2007, Rudick et al. 2010). However, accurate photometric measurements of the
diffuse light are difficult to perform because its surface brightness is typically 
fainter than
1\% of that of the night sky, and it can be difficult (especially at high redshift) to
distinguish the extended outer halos of the brightest cluster galaxies (BCGs) in a
cluster core from the stars floating freely in the cluster potential, i.e. the
intracluster light component. Moreover, the dimming factor plays an important role in
our study and makes it intrinsically difficult to detect ICL at high redshift. This
cosmological surface-brightness dimming, which follows a $(1+z)^4$ law, places {\it de
facto} an upper limit on the redshift to which the ICL can be practically studied,
that depends on the image depth and the detection methods.

This explains why, until now, most studies of the ICL have been
performed on galaxy clusters at redshifts below z$\sim$0.3 (see
e.g. Toledo et al. 2011).  However, since it is crucial to understand
how the evolution of galaxy clusters affects that of the ICL, we must
study the ICL within a range of clusters at various redshifts. It
would be ideal to investigate as much as possible the period between
z=0.3 and z$\sim$2, and investigate clusters since their birth. We
propose here to fill part of this gap in the 0.4-0.8 redshift range
for a sample of ten clusters.  This redshift range is sufficient to
cover about half of the typical cluster lifetimes. To help us follow the
cluster formation history, we also compare our ICL results in terms of
cluster prefered orientations obtained here with those obtained with
the Serna $\&$ Gerbal (1996) dynamical method.
 
Colors are useful for determining the evolution of galaxy clusters,
and we have been able to 
 detect the ICL in two bands (the HST/ACS F814W-band and the ground-based VLT/FORS2
V-band) for the three lowest redshift clusters. In summary, we studied the diffuse light in
ten different clusters in one band up to z$\sim$0.8 based on deep HST ACS images and in three
of them (up to z$\sim$0.58) in two bands with FORS2 data.
However, these deep data would not have been sufficient if we were not using a very sensitive
wavelet-detection technique, the ov\_wav method
(Pereira 2003 and Da Rocha et al. 2005), itself a variant of the \`a trous wavelet transform
described by Starck et al. (1998, see also Starck et al. 2002). This method is independent of
both the galaxy and star modelling, and of the sky level subtraction.

In Sect.~2, we describe the sample and observational data. In Sect.~3, we present the data
analysis methods and detection efficiency estimates. In Sect.~4, we present our analysis for
the F814W images. In Sect.~5, we discuss the nature of our detections, taking advantage of our
V-band data. Finally, the main results are summarized in Sect.~6. Throughout the paper, we
assume that H$_0$ = 71 km s$^{-1}$ Mpc$^{-1}$, $\Omega _m$=0.27, and $\Omega _{\Lambda}$=0.73.
All magnitudes are in the $AB$ system.

\section{Sample and observational data}

\subsection{The cluster sample}

Our study is focused on ten clusters from the Las Campanas Distant
Cluster Survey (hereafter LCDCS: Gonz\'alez et al. 2001) observed in
the HST/ACS F814W-band, and for three of these ten clusters also in
the VLT/FORS2 V-band. The LCDCS sample consists of optically selected
clusters in the z=[0.3, 1] range identified from fluctuations in the
extragalactic background light.  The LCDCS was a drift-scan imaging
survey of a 130~deg$^2$ strip of the southern sky. Selection criteria
for the LCDCS survey are fully automated so it constitutes a
well-defined homogeneous sample that can be used to address issues of
cluster evolution and cosmology. Several studies have already been
performed for these individual clusters (e.g. Halliday et al. 2004 and
Milvang-Jensen et al. 2008).

The ten considered clusters are also part of the DAFT/FADA Survey,
which consists of 91 distant clusters ranging from z=0.4 to z=0.9 and for
which we have images in several bands \footnote{(see http://cencos.oamp.fr/DAFT/ and
Guennou et al. 2010 for details)}.

These ten clusters were chosen because of the very deep HST/ACS F814W
data available (prop. 9476, PI: J. Dalcanton), especially in a central
$\sim 2 \times 2$~arcmin$^2$ area (see next section). We present in
Table~\ref{sample} the basic characteristics of these ten clusters,
including the redshift, velocity dispersion, and mass-to-light ratio
(M/L) from Clowe et al. (2006), and the total ICL F814W absolute
magnitude.

\begin{table*}
\caption{Main characteristics of the 10 clusters observed, and total ICL F814W 
absolute magnitudes. The numbers between parentheses
in the redshift column are the number of literature redshifts used for the Serna $\&$
Gerbal analysis.}
    \center
\begin{tabular}{ccllc}
\hline
\hline
Name & redshift (nb) & velocity dispersion & cluster M/L ratio & ICL total\\
     &          & (km/s)              & (M$_\odot$ / L$_\odot$) & F814W mag.\\
\hline
LCDCS 0110 & 0.580 (111) & ~319 (+53 -52)  & 183 (+66 -67) & -20.1\\
LCDCS 0130 & 0.704 (95) & ~418 (+55 -46) & ~74 (+54 -56) & -19.8\\
LCDCS 0172 & 0.697 (84) & ~589 (+78 -70)  & 151 (+36 -37) & -20.8\\
LCDCS 0173 & 0.750 (83) & ~504 (+113 -65)  & 217 (+44 -46) & -20.2\\
LCDCS 0340 & 0.480 (81) & ~732 (+72 -76)   & 209 (+66 -69) & -17.9\\
LCDCS 0504 & 0.794 (97) & ~1018 (+73 -77)  & 130 (+16 -17) & -21.0\\
LCDCS 0531 & 0.634 (93) & ~574 (+72 -75) & ~65 (+128 -65) & -19.1\\
LCDCS 0541 & 0.542 (72) & ~1080 (+119 -89)   & 247 (+27 -28) & -20.4\\
LCDCS 0853 & 0.757 (89) & ~648 (+105 -110)  & 160 (+39 -39) & -20.1\\
CL J1103.7-1245a & 0.640 (89) & ~336 (+36 -40)  &  & -18.8\\
\hline
\end{tabular}
\label{sample}
\end{table*}

\subsection{Observational data}

\subsubsection{HST ACS data}

We have at our disposal HST ACS F814W data, each image including 4
tiles (2$\times$2 mosaic) of 2~ks and a central tile of 8~ks (Desai et
al. 2007). The depth achieved for point sources at the 90$\%$ level is
of the order of 28 magnitude for the deep parts and 26 magnitude for
the shallow parts in F814W (see Guennou et al. 2010).  These values
are only estimates of the true completeness levels as the simulated
objects do not perfectly reproduce the full range of real object
profiles. In addition to the different exposure times, the different
thresholds used in SExtractor for the shallow and deep regions also
explain the two magnitude difference between the two completeness
levels\footnote{This is based on total magnitudes and an analysis
  performed at the 1.8$\sigma$ and 3$\sigma$ SExtractor levels,
  respectively, for the deep and shallow parts to limit fake object
  detections.}. The full data reduction technique is described in
Schrabback et al. (2010) but here we describe the salient points. The
data were reduced using a modified version of the HAGGLeS pipeline,
with careful background subtraction, improved bad pixel masking, and a
proper image registration. Stacking and cosmic ray rejection were done
with Multidrizzle (Koekemoer et al. 2002), taking the time-dependent
field-distortion model from Anderson (2007) into account. The pixel
scale was 0.05 arcsec and we used a Lanczos3 kernel. After aligning
the exposures of each tile separately, we determined the shifts and
rotations between the tiles from separate stacks by measuring the
positions of objects in the overlap regions. As a final step, we
created mosaic stacks by including all tiles for a given cluster.

\subsubsection{FORS2 V-band data}

FORS2 is a multi-mode (including an imaging-mode) optical instrument
mounted on the ESO VLT/UT1 Cassegrain focus. The considered images
had 0.25~arcsec pixels, covering a 6.8$\times$6.8~arcmin$^2$ field of
view. Guennou et al. (2010) showed that the V-band images we selected
were basically complete down to F814W$\sim$26 for point sources. We
refer the reader to White et al. (2005) for further details about the
observations.

\section{Methods}
 
\subsection{Wavelet analysis}
 
Deep images are not always sufficient to detect ICL in distant
clusters because of the dimming factor. We therefore applied the very
sensitive ov\_wav package (see e.g. Pereira 2003 and Da Rocha \&
Mendes de Oliveira 2005) to HST F814W images and FORS2 V-band
images. The method is a multi-scale vision model (e.g. Ru\'e and
Bijaoui 1997), and we now describe its main steps. After applying the
wavelet transform (i.e. the \`a trous algorithm) to a given ``sky
image" (direct observation of the targeted cluster) a thresholding in
the wavelet transform space is performed to identify the statistically
significant pixels. These are grouped in connected fields by a scale
by scale segmentation procedure, in order to define the object
structures. An inter-scale connectivity tree is then established and
the object identification procedure extracts each connected tree that
contains connected fields of significant pixels across three or more
scales and, by referring to the object definition, can be associated
with the objects.  Finally, an individual image can be recovered for
each identified object, by applying an iterative reconstruction
algorithm to the data contained in the corresponding tree. Both
measurement and classification operations can then be carried out.

Since the images were by far too wide (8k$\times$8k) for the code to operate
efficiently, we divided the total area to be analyzed into five subareas 
(see below).

In a first step, we detected high frequency objects in a two pass 
process. These objects were first detected in the $sky$ image to produce 
our first $object$ image.  We used characteristic
scales between 1 and 1024 pixels in the wavelet space. The first $object$ 
image was then subtracted from the $sky$ image to
produce the first $residual$ image. 
 This first $residual$ image therefore still includes objects larger than 
1024 pixels (51.2 arcsec diameter) as well as several smaller previously hidden 
features. These hidden features are typically  objects too faint to satisfy
the wavelet first pass thresholding conditions described in the following. 
We then detected objects in this
$residual$ image, with characteristic
scales between 1 and 128 pixels,
to create a second $object$ image still including objects larger than 
128 pixels (6.4 arcsec diameter) as well as several smaller previously hidden 
features. This second $object$ image was then
subtracted from the original $sky$ image to produce a second
$residual$ image.  The main parameter for the object detection in the
wavelet method was a significance level of at least 5$\sigma$ for each
of the two passes. 

At this point, most of the major compact objects have been removed
from the second $residual$ image. Small scale (radius smaller than 3.2 arcsec) 
bright diffuse light sources also have been removed. At z$\sim$0.6, this means 
that plume-like structures (see e.g. Gregg $\&$ West 1998) of less than 
$\sim$40 kpc have been removed from the second $residual$ image. In a similar way 
as in Adami et al. (2005), the third pass involves a search for what we 
will call the real ICL sources, i.e. the most extended surface brightness 
features in the second $residual$ images. 
These features were detected by considering pixels where the signal was
larger than 2.5$\sigma$ (for each of the clusters), compared to
an empty area of the $residual$ image. Each of these
sources was visually inspected before the removal of obvious residuals
of bright saturated Galactic stars or defects due to image cosmetics
(for example areas of low signal-to-noise ratio). By definition,
  ICL sources  do not include star residuals. 

We show in Fig.~\ref{ex} the results  for the central area of
LCDCS~0541. Objects detected in the first two passes are all
compact and are similar to classical galaxies or stars. A large part
of them also are detected by a SExtractor analysis, with
the exception of several very faint objects that we will later discuss
and refer to as ``wavelet detected compact objects'' (hereafter WDCOs).
These WDCOs are  wavelet-detected (during the two first passes) 
objects which would not have been detected by SExtractor.

Objects detected in the third pass are several tens of kpc wide; they
are not compact objects such as galaxies, but are what we refer to as real
ICL sources.

\begin{figure}[!h]
\begin{center}
\caption{1.1$\times$1.1~arcmin$^2$ images of the various steps of the
  wavelet analysis for LCDCS~0541. Upper left: $sky$ image, upper
  right: first pass objects, lower left: second pass objects, lower
  right: final $residual$ image.}
\label{ex}
\end{center}
\end{figure}

\begin{figure}[!h]
\begin{center}
\includegraphics[angle=270,width=3.00in]{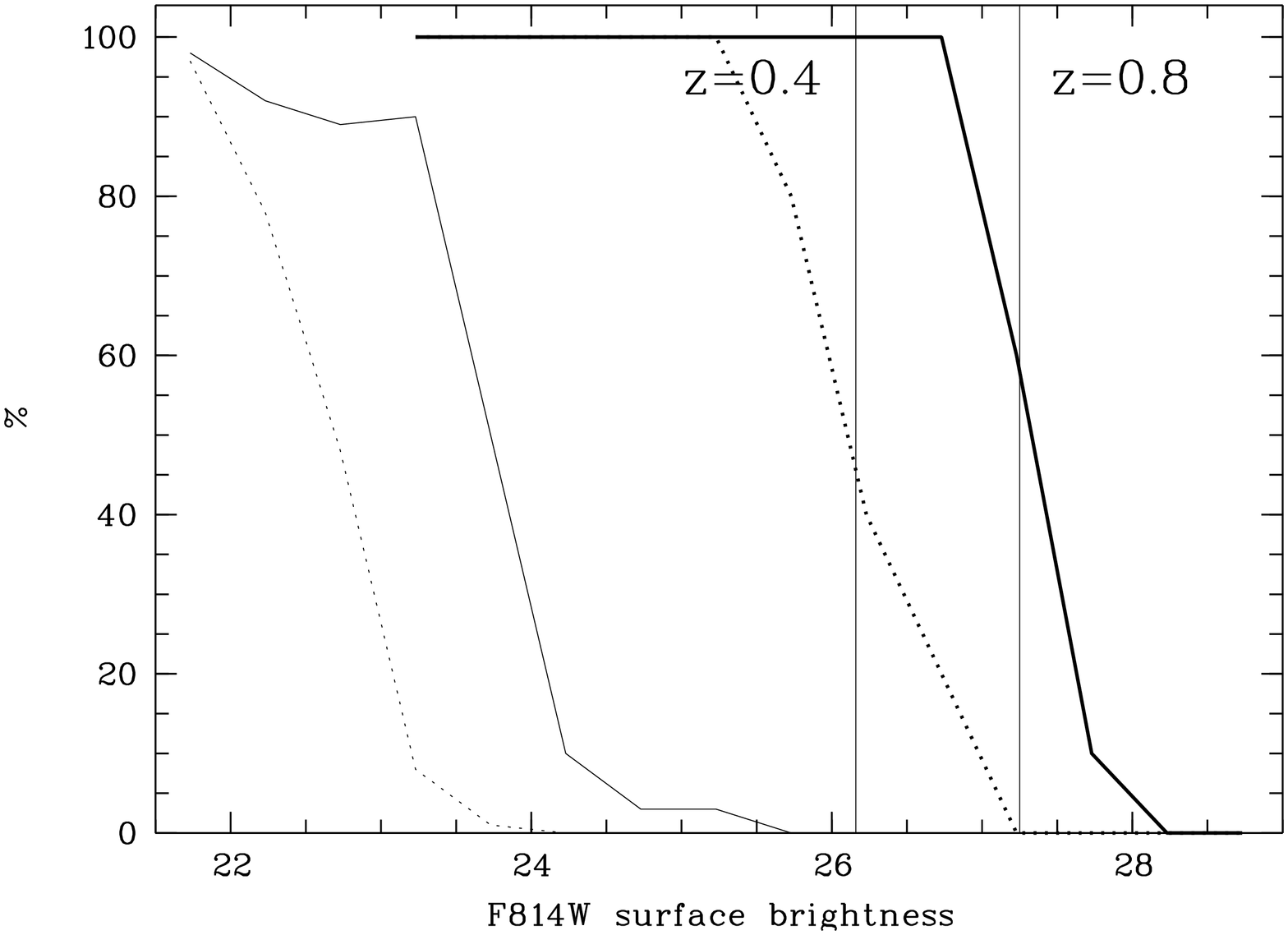}
\includegraphics[angle=270,width=3.00in]{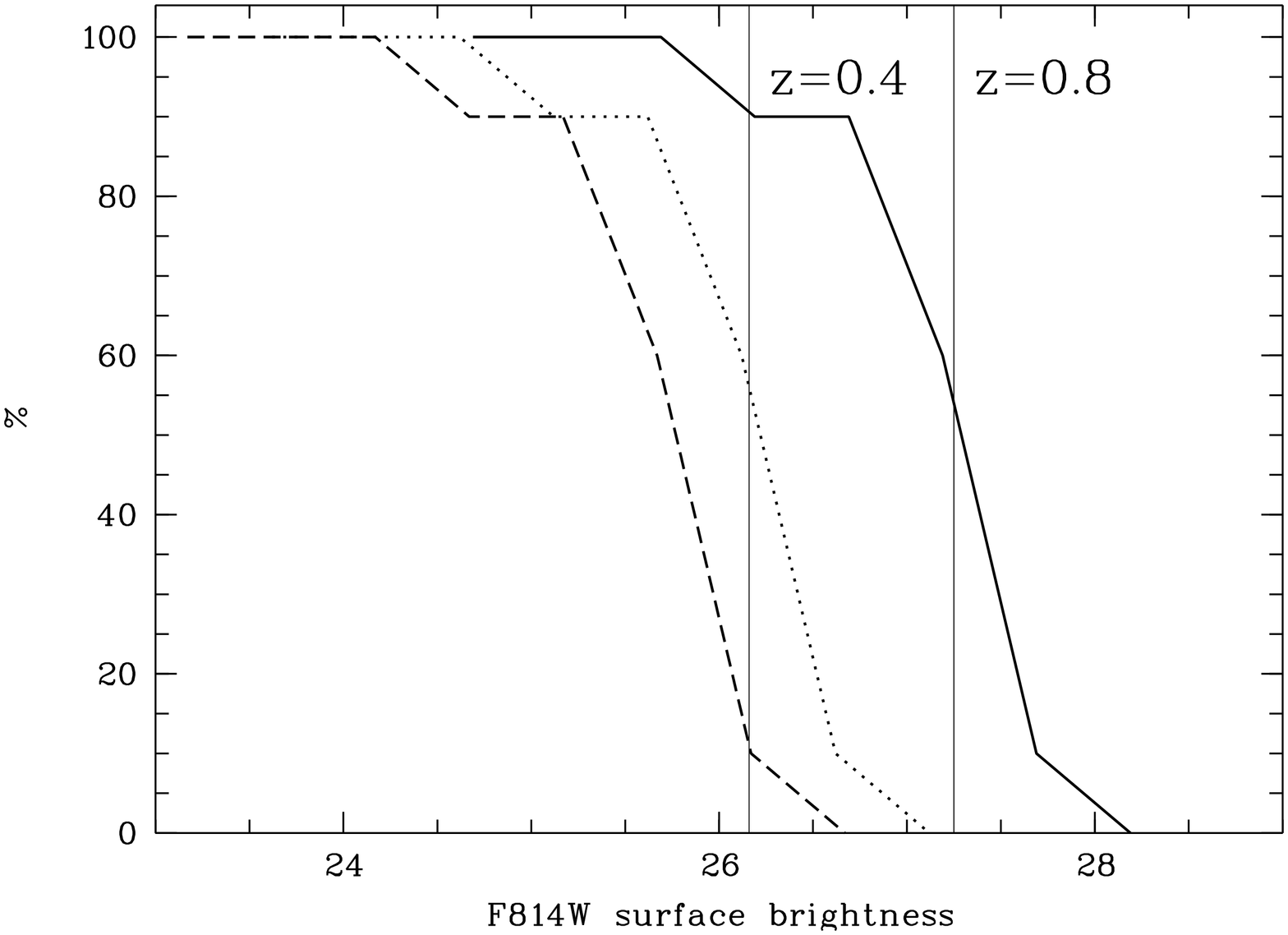}
\caption{Upper panel: percentage of recovered diffuse light sources
  in HST ACS images. Curves to the right are related to the wavelet
  detections. Curves to the left are relative to the SExtractor
  detections. Dotted curves are relative to shallow HST ACS image
  areas. Continuous curves are relative to deep HST ACS image
  areas. The two vertical lines show the brightest Coma cluster known
  diffuse light source (a $\sim$60 kpc wide source) redshifted to z=0.4 and z=0.8. Lower panel:
  percentage of recovered diffuse light sources in FORS2 V-band
  images. The two vertical lines show the brightest known Coma cluster
  diffuse light source redshifted to z=0.4 and z=0.8.
  All curves correspond to the wavelet detection. The continuous
  curve represents very blue objects (V-F814W=1.), the dotted 
curve blue objects (V-F814W=2.1),
  and the dashed curve red objects (V-F814W=2.6). }
\label{det}
\end{center}
\end{figure}

\begin{figure}[!h]
\begin{center}
\includegraphics[angle=270,width=3.00in]{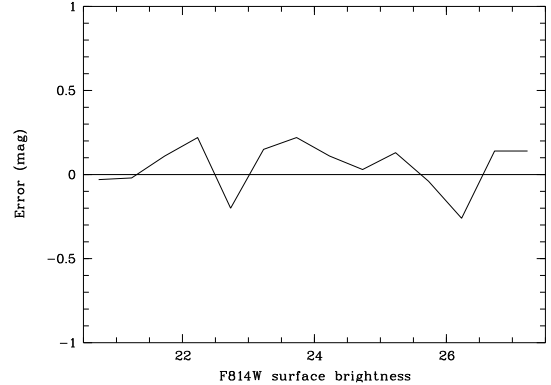}
\caption{Mean error in the surface brightness estimate as a function
  of the F814W surface brightness.}
\label{magt}
\end{center}
\end{figure}

\begin{figure}[!h]
\begin{center}
\caption{Upper panel: F814W HST ACS image of the central area of LCDCS~0541, showing
  the third wavelet pass residuals. White contours correspond to
  levels starting at 2.5$\sigma$ and increasing by 0.5$\sigma$.  Blue
  ellipses show the classical SExtractor-detected objects (not visible
  since they are removed by the first two wavelet passes). The image
  is 1.6$\times$1.6~arcmin$^2$ wide. Lower panel: same as before using only the 1 orbit exposures. This image is zoomed on the central ICL source. The first white contour is at the 1.5$\sigma$ level.}
\label{ex1232}
\end{center}
\end{figure}

\begin{figure}[!h]
\begin{center}
\includegraphics[angle=0,width=2.70in]{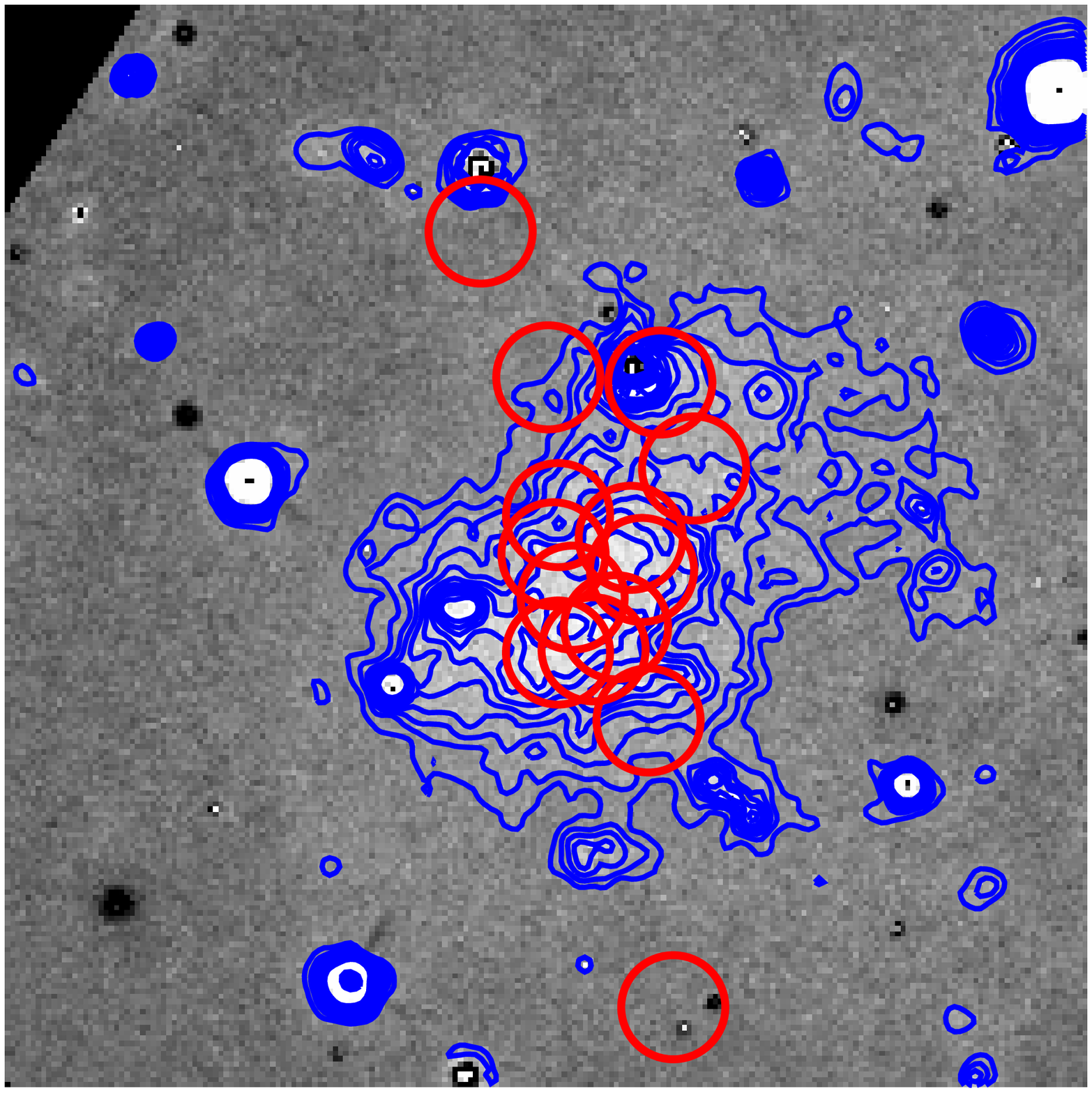}
\includegraphics[angle=0,width=2.70in]{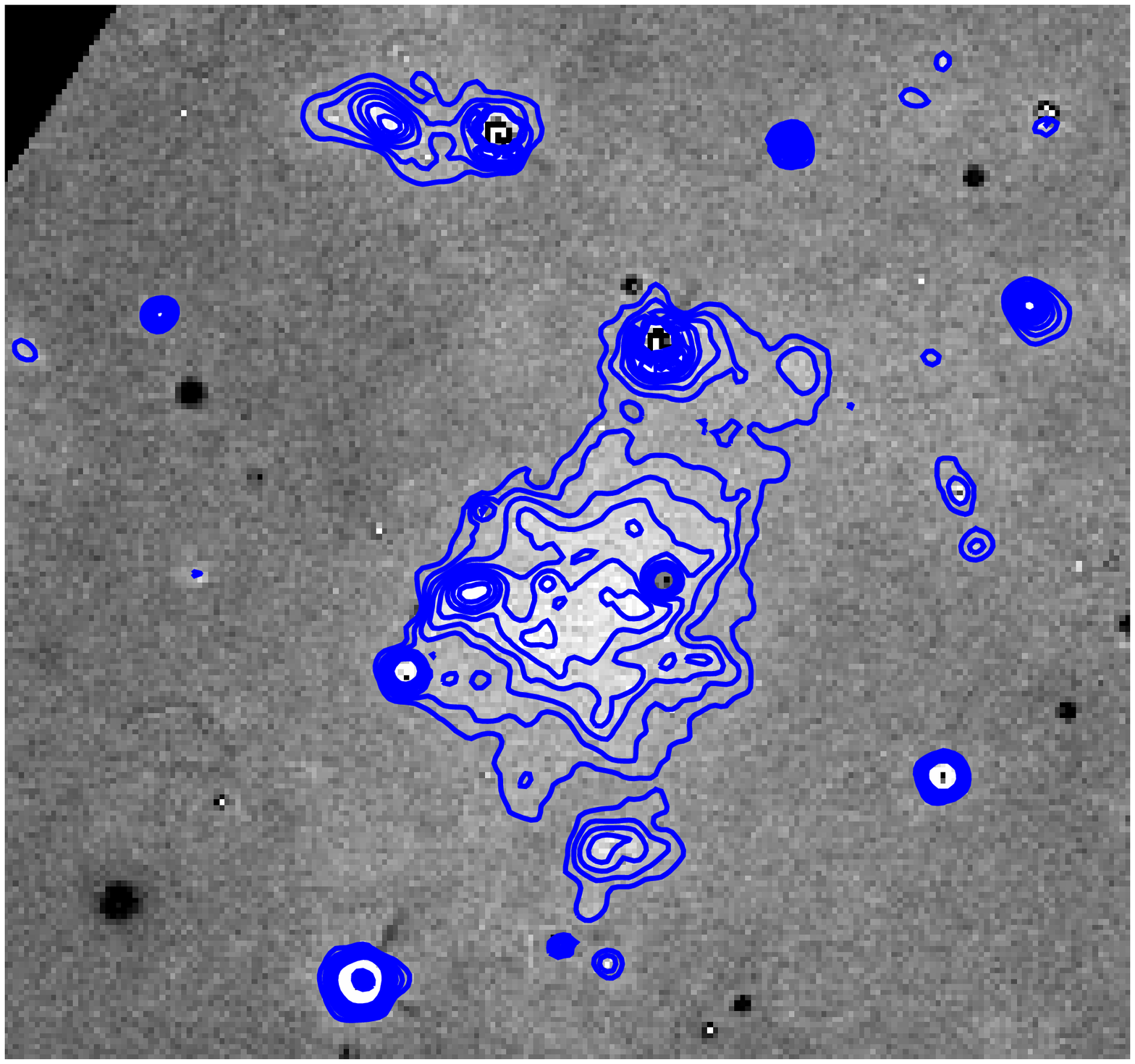}
\includegraphics[angle=0,width=2.70in]{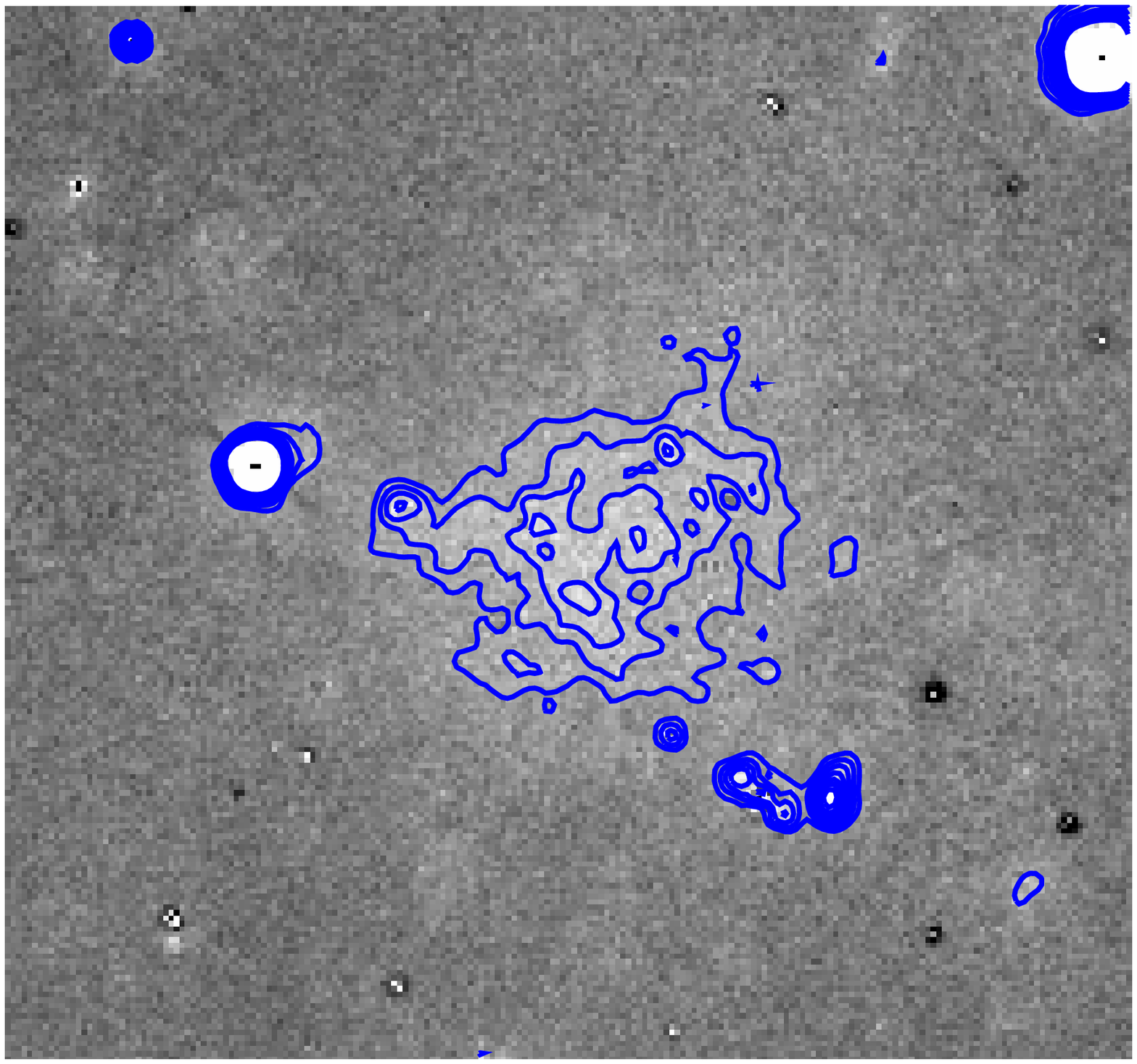}
\caption{Stacked F814W HST ACS second $residual$ images of all ten
  considered clusters (upper figure), of the z$\leq$0.65 clusters
  (middle figure: 5 clusters taken into account), and of the z$\geq$0.70 
  clusters (lower figure: 5 clusters taken into account). Images are 
  200$\times$200~kpc$^2$ wide. The upper figure
  also shows as red circles the positions of the brightest cluster
  galaxies of the stacked clusters.}
\label{stack}
\end{center}
\end{figure}

\subsection{Detection efficiency}

Before attempting any analysis of our detections, a crucial step is to
assess the detection levels of our method as a function of the surface
brightness of the diffuse light sources. No diffuse light source
catalog is available for the considered cluster sample, so we rely
entirely on simulations to assess our ability to detect these
sources. To achieve this, we employed a classical method where
artificial sources are introduced into real images and then detected
with our codes. The same method was applied for example in Adami
et al. (2006) to estimate the detection levels of low surface brightness
galaxies in the Coma cluster. Since both HST ACS and FORS2 V-band
exposure times and observing conditions were similar for all the
considered clusters (see e.g. Guennou et al. 2010), we arbitrarily
selected the LCDCS~0541 cluster F814W and V images to perform our tests.

As a model of diffuse light sources, we chose a uniform disk with a radius
of 25 pixels that we scaled to different surface brightnesses. These 25
pixels correspond to a $\sim$20~kpc wide object at z$\sim$0.8, which is
typical of major-galaxy sizes. It is an intermediate size between the 
WDCOs and the real ICL sources. The predicted detection levels will
therefore be intermediate between the WDCOs and the true detection 
rates of the ICL sources. Surface brightnesses were chosen in successive
0.5 mag wide bins between F814W = 21.5 and 28.5 mag/arcsec$^2$. We used
various V-F814W colors to include these objects in the V-band images, namely
V-F814W=1., 2.1, and 2.6. These values are typical of very blue, blue, and red
objects (see Fig.~\ref{det}).

For each magnitude bin, we repeated the exercise ten times to compute
a mean detection rate. We were unable to repeat this 100 times as in
Adami et al. (2006) because the wavelet detection algorithm would have
required a prohibitively long computing time. The whole process was performed
in various places in both the deep (10 ks) and the wide (2~ks) HST ACS
image areas. In Fig.~\ref{det}, we display our results and compare them with the
detection levels achieved using a classical SExtractor detection
(Bertin $\&$ Arnouts 1996, with a $1.8\sigma$ detection threshold and a minimum 
significant area of 4 pixels) instead of the wavelet technique.

We clearly see that SExtractor detections are basically impossible for
objects fainter than F814W$\sim$24 mag/arcsec$^2$, even when considering the
deepest parts of the HST ACS images. The use of the wavelet technique allows a
magnitude gain of at least 3.5 (at a cost of a factor of nearly 1000 in computing time).
 
We roughly estimated whether these detection levels are sufficient to
detect typical diffuse light sources redshifted to the considered
cluster distances. We selected the brightest diffuse light
source detected in Coma (source 3 of Adami et al 2005) and estimated its F814W 
surface brightness
by using the colors given in Fukugita et al. (1995) and applying (1+z)$^4$
cosmological dimming. This source is very similar to the ICL sources listed in 
Table~\ref{sample}, with a diameter of $\sim$60 kpc, no significantly peaked surface
brightness profile, and a mean surface brightness of 
R$_{Vega}$$\sim$25 at z=0.023. The deduced F814W surface brightnesses at z=0.4 and 
z=0.8 (extreme redshift values of our cluster sample) are shown in Fig.~\ref{det} as 
the two vertical lines. However, we must keep in mind that these vertical lines are
valid for very extended ICL sources while the  detection levels shown in Fig.~\ref{det}
are for smaller (25 arcsec radius) objects. It appears that the shallow
parts of the HST ACS images are insufficiently deep to allow diffuse light
source detections for a large part of our cluster sample, even when using the wavelet 
detection technique. We therefore consider only the deepest parts of the HST ACS images.

For the FORS2 V-band images, Fig.~\ref{det} shows that we can only
expect to detect blue to very blue objects and not at redshifts
significantly higher than 0.4. We therefore chose to consider only the
three clusters in the sample at z$\leq$0.58 to attempt the detection
of diffuse light sources (WDCOs) in the V-band.

Finally, we verified our ability to recover the true surface
brightness value after applying the wavelet detection
process. Fig.~\ref{magt} shows the mean error in the surface
brightness estimate as a function of surface brightness. This error
was calculated to be the mean difference between the chosen surface
brightness of the artificial diffuse light sources and the measured
one. We see that our estimated uncertainty is smaller than 0.25
magnitudes over the entire magnitude range considered. This value
  is taken to be the typical magnitude uncertainty in the following.

\section{F814W diffuse light detections}

\subsection{Stacked images}

The first two passes of the wavelet process essentially provide detections
of faint compact objects, the WDCOs. These objects are
not diffuse light sources but rather compact galaxies (or stars) that were
too faint to be detected by SExtractor. In Sect.~5.2, we discuss the WDCOs
that were detected in both the F814W and the V-band images.

Here, we concentrate on the third pass objects, which are much larger
in size and represent the real ICL sources. These objects are very similar to
those detected in Coma (Adami et al. 2005) and A2667 (Covone et
al. 2006a), and extend over several tens of kpc. In Fig.~\ref{ex1232}, we show
examples of these objects detected at the 3$\sigma$ level at the LCDCS~0541
center. None of these objects (when obvious residuals from bright saturated
Galactic stars are removed) can be stars.

To establish whether all of the detected large-scale diffuse light sources are
located at the cluster centers, we applied a stacking technique similar to the
one employed by Zibetti et al. (2005). We rescaled all the second 
$residual$ images in units of kpc and added them together using the EDISCS 
cluster centers as reference nodes. Results are shown in Fig.~\ref{stack}.

These diffuse light sources have typical sizes of a few tens of
kpc. This implicitly assumes that the detected diffuse light sources are
at the cluster redshifts, a hypothesis which does not seem unreasonable
given the characteristics of these sources. Fig.~\ref{stack} shows a
very clear 8$\sigma$ detection (at the source center) extending over a
$\sim 50\times 50$~kpc$^2$ area. The total absolute magnitude of this
source is $-21.6$ in the F814W filter, which is equivalent to about two
$L^*$ galaxies for each of the ten clusters (assuming the $L^*$ values of
Ilbert et al. 2006). 

Total absolute magnitudes were measured first evaluating the apparent
magnitude of the diffuse light sources. This was done by measuring the
total flux in the second $residual$ image inside a circle immediately
surrounding the considered source (limited by the 2.5$\sigma$ external
coutour, typically slightly more than 200 pixels). Background was
estimated in a of 10 pixel (5 arcsec) wide concentric annulus
surrounding the measurement circle and starting 50 pixels after the
measurement circle radius. Star residuals were not included in the computation 
in this annulus.

Then, as previously, we assumed the diffuse light source to be part of
the considered cluster in order to compute a distance modulus with the
adopted cosmology.  Finally, we k-corrected the computed absolute
magnitudes with the values listed in Section 4.5., assuming
elliptical-like colors.

Fig.~\ref{stack} also shows the brightest cluster
galaxies of the stacked clusters. This allows us to see that the detected
diffuse light sources are most of the time close to these galaxies, but not
always centered on these positions.

We divided our sample into low and high redshift clusters and produced
two stacked $residual$ images for clusters at z$\leq$0.65 and
z$\geq$0.70.  This permitted 6$\sigma$ and 8$\sigma$ detections in
both redshift bins (see Fig.~\ref{stack}). The absolute magnitudes of
the diffuse light sources detected at z$\leq$0.65 and z$\geq$0.70
($-20.9$ and $-21.3$, respectively, in the F814W filter) correspond to
about two $L^*$ galaxies for each of the individual clusters.  It is
not easy, however, to express the amount of ICL as a percentage of the
total cluster light, because it requires a measurement of the total
cluster light. In the considered redshift range, we were only able to
use photometric redshifts to define cluster membership, and the
relatively large photometric redshift uncertainties ($\pm$0.1: see
Guennou et al. 2010) would have in turn created too large an
uncertainty on the measurement of the total cluster light.

We note that in the LCDCS detection technique the clusters are
detected as positive surface brightness fluctuations in the background
sky.  If the amount of  diffuse light detected was greater than the
galaxy contribution, then the detection method would preferentially
identify clusters with a large amount of diffuse light. Given the
moderate amount of diffuse light, this bias, even if present, should
not have a strong effect.

\subsection{Superflat residuals?}

In this subsection, we check by two ways that our ICL detections are
not artefacts due to the numerical image reduction. Superflat defaults
may produce such sources.

A good method to check this is to try to redetect our ICL sources in HST
F814W ACS images where they physically lie in different CCD
regions. This would be possible by using only the external 1 orbit
tiles in order to generate the central image. This means that we would
exclude the central 4 orbits tile and that the area where the ICL
source is present will not be affected in the same way by the possible
numerical artefacts. The problem is that we showed that 1 orbit images
are not deep enough to efficiently detect ICL sources in the most
distant of our clusters. We therefore only selected LCDCS~0541, which
is one of the closest structures in our sample (z=0.54) and the
structure with the second brightest detected ICL source. We analyzed
for this cluster the 1 orbit images in the same way as the full
depth image. We show the result in Fig.~\ref{ex1232}. We redetect the
ICL source, as expected at a lower signal to noise of 1.5. The
magnitude of this source is $-19.6$, fainter than the $-20.4$ value
predicted by the full depth image. This is not surprising since we do
not detect the external part of the ICL source with images taken
during only one orbit.

Another test is to search for diffuse light in places where it is not
supposed to be present. For this, we selected all the central areas
(1000$\times$1000 pixels) of the external 1 orbit tiles. These areas
only cover the cluster outskirts where we did not detect any
significant ICL sources. We then added together these images
disregarding their astrometry and produced in this way a very deep
fake image supposed to be free of ICL sources. This image is made of
forty 1 orbit tiles and provides an image quality similar to that of
the full depth central images. Analyzing this fake field in the same
way as previously, we produced a $residual$ map which shows nearly no
significant ICL source (see Fig.~\ref{nulltest}). Numerical artefacts
therefore do not seem to be a problem in our analysis.

\begin{figure}[!h]
\begin{center}
\caption{Fake image built with the forty external 1 orbit tiles. We
  only show the area of 1000$\times$1000~pixel$^2$. Red contours
  correspond to levels starting at 2.5$\sigma$ and increasing by
  0.5$\sigma$. The large black square represents the typical size of
  ICL sources detected in real images. The black circle is the only
  diffuse light source detected in this fake image which is not a star
  residual.}
\label{nulltest}
\end{center}
\end{figure}

\subsection{Anisotropic ICL distribution?}

The stacked images presented in the previous section were obtained by
assuming random orientations of the clusters. The images of the ICL
were therefore, by definition, isotropic.  To determine whether any
preferential orientation could be identified in the spatial distribution
of the ICL, we chose to rely on another measurement of the
orientation, based on the substructures potentially populating the
clusters (we could also have relied on the brightest cluster galaxy
orientation but the goal of the present paper is not a detailed study
of these galaxy characteristics). As shown in several studies
(e.g. Adami et al. 2009), these substructures are good indicators of
the directions in which matter is being accreted from the cosmic web,
and when they are detected they appear as special orientations of the
clusters.

\begin{figure*}[!h]
  \centerline{
  \mbox{\includegraphics[angle=270,width=3.0in]{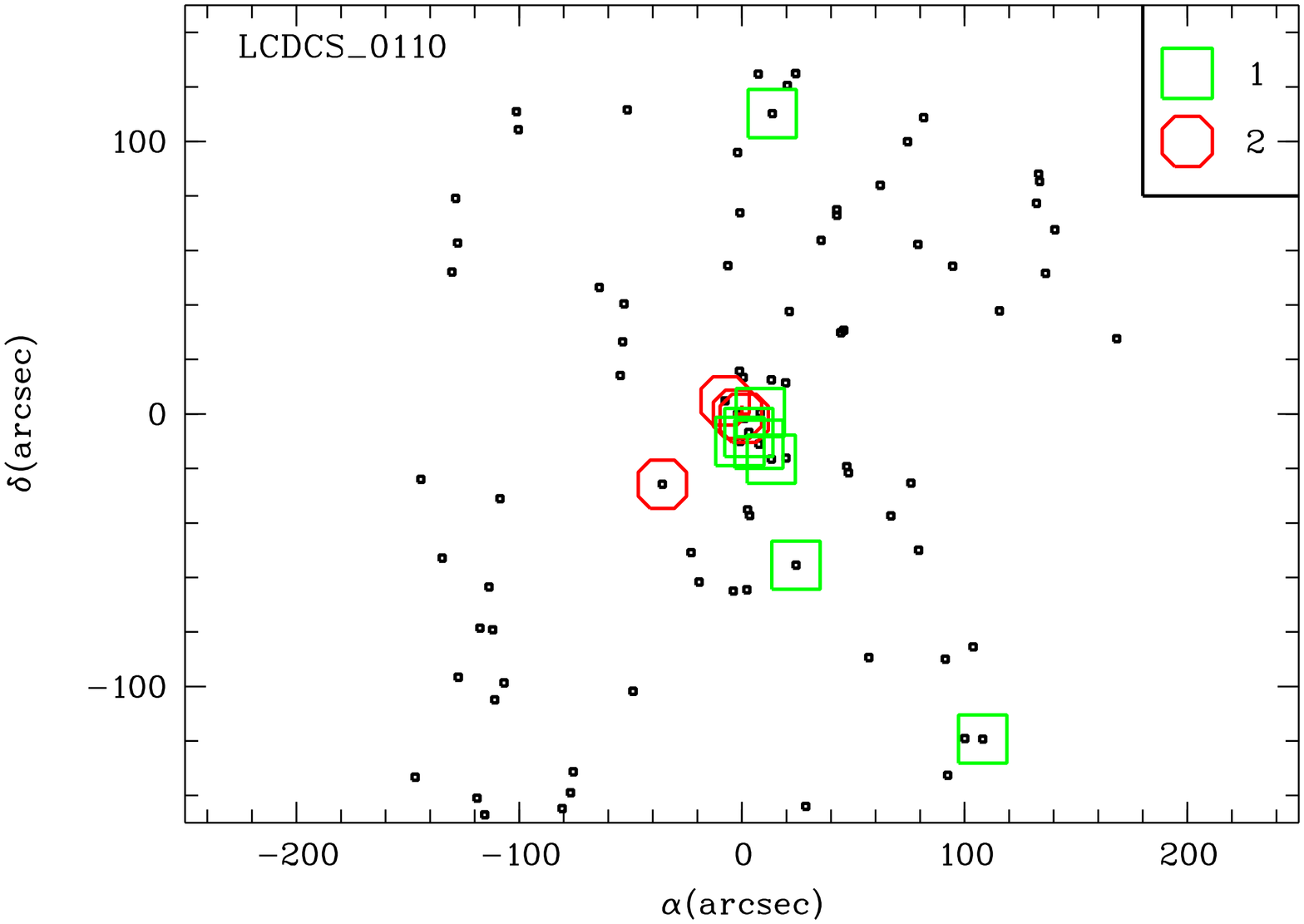}}
  \mbox{\includegraphics[angle=270,width=3.0in]{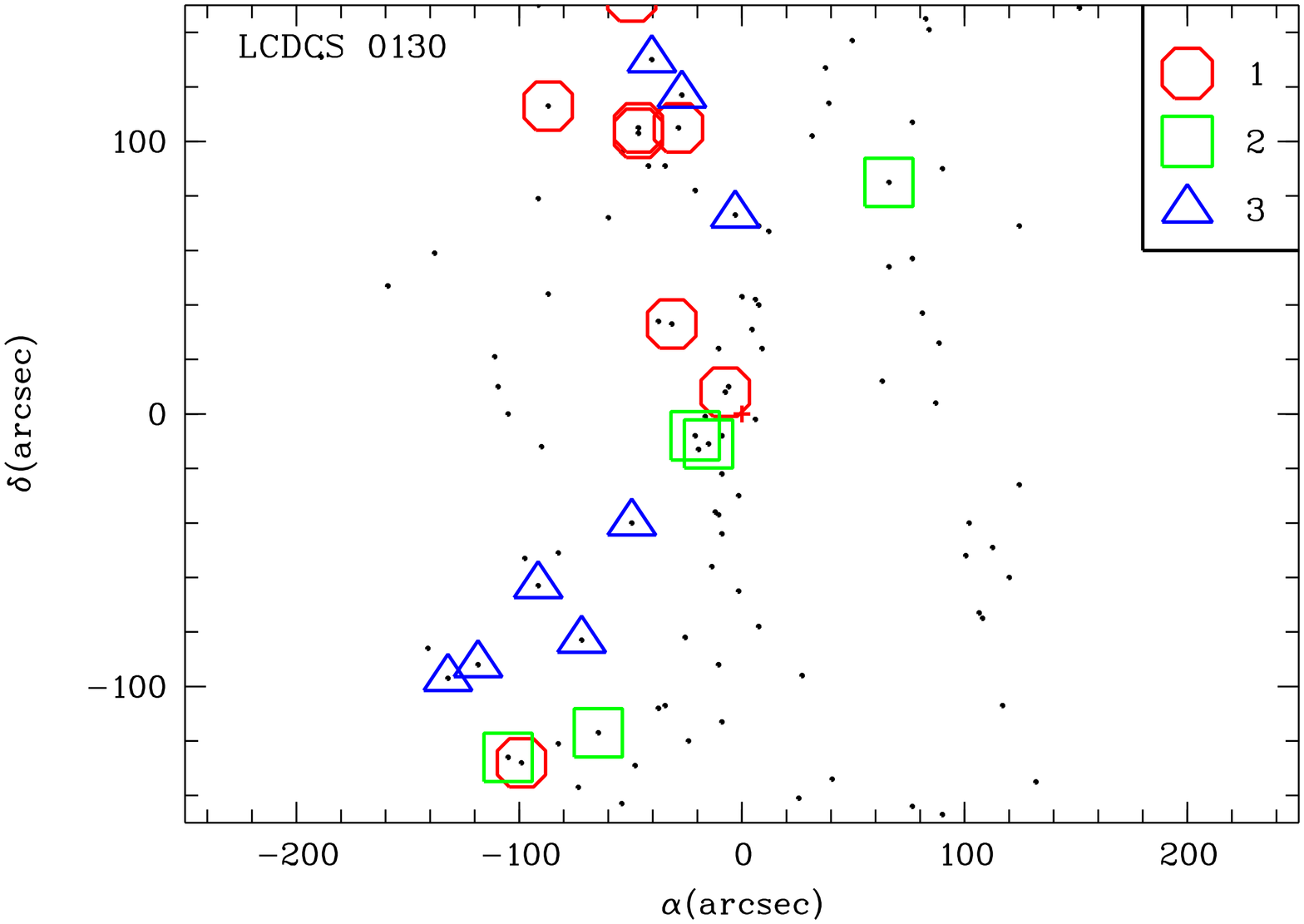}}
  }
  \centerline{
  \mbox{\includegraphics[angle=270,width=3.0in]{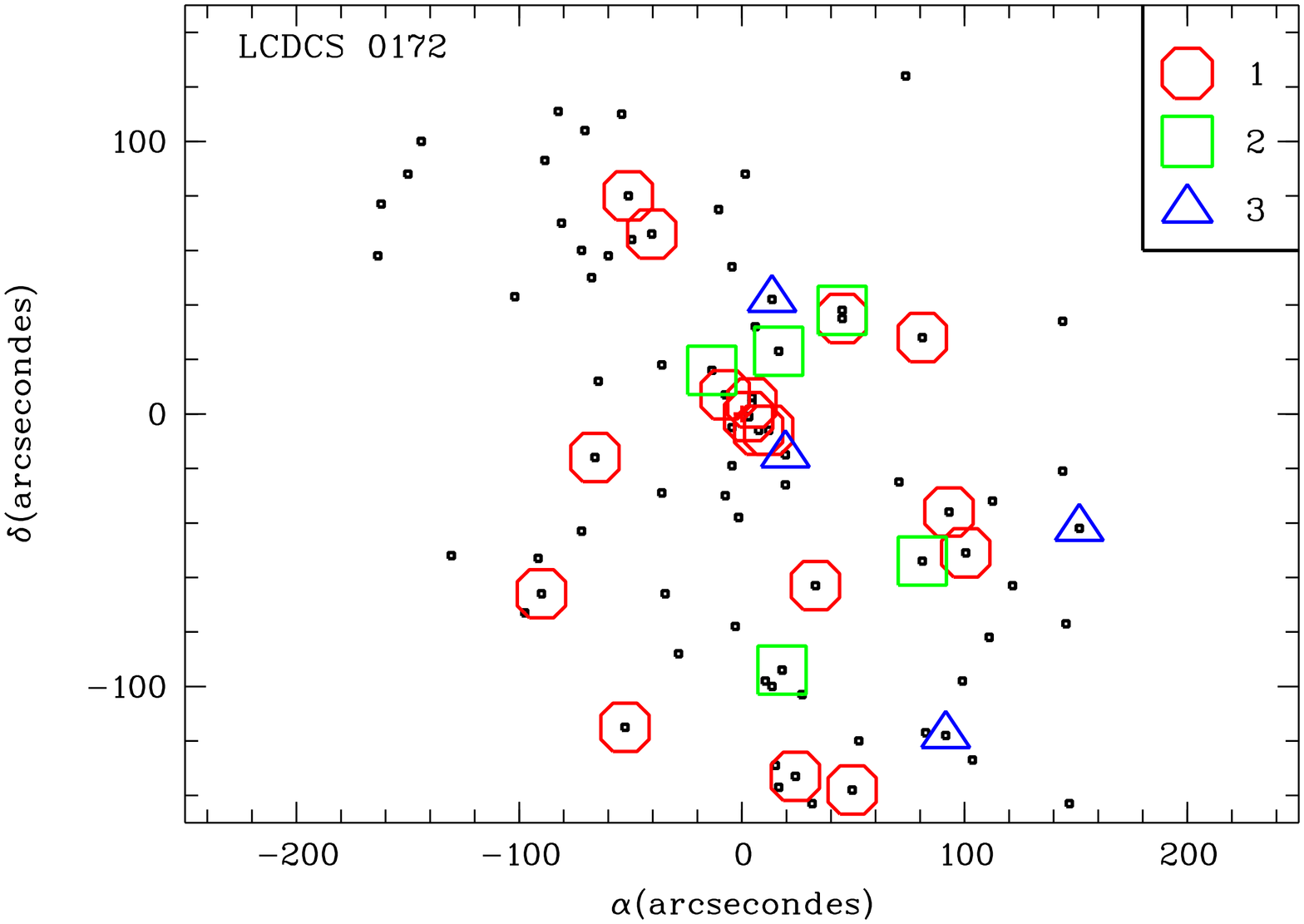}}
  \mbox{\includegraphics[angle=270,width=3.0in]{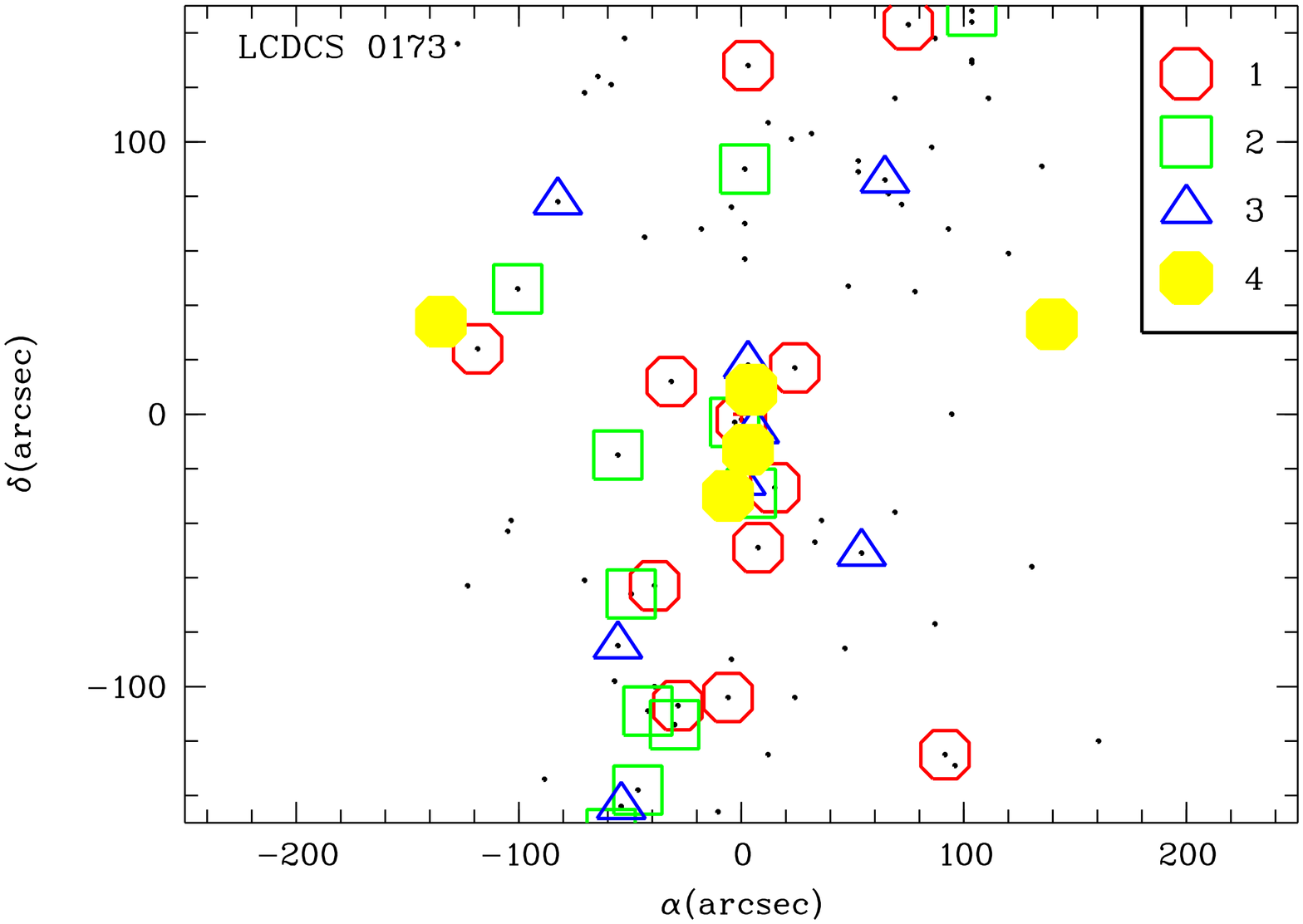}}
  }
  \centerline{
  \mbox{\includegraphics[angle=270,width=3.0in]{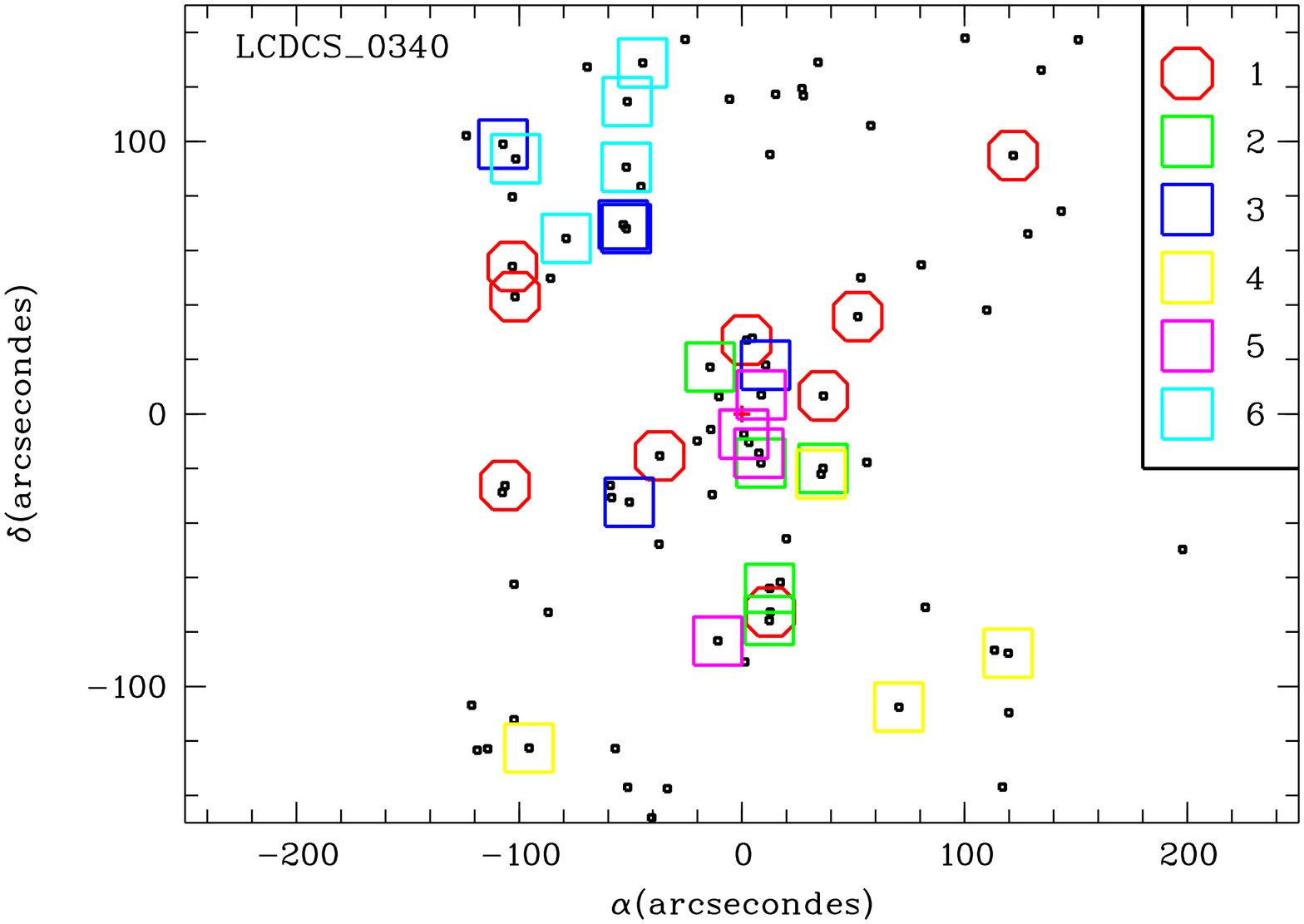}}
  \mbox{\includegraphics[angle=270,width=3.0in]{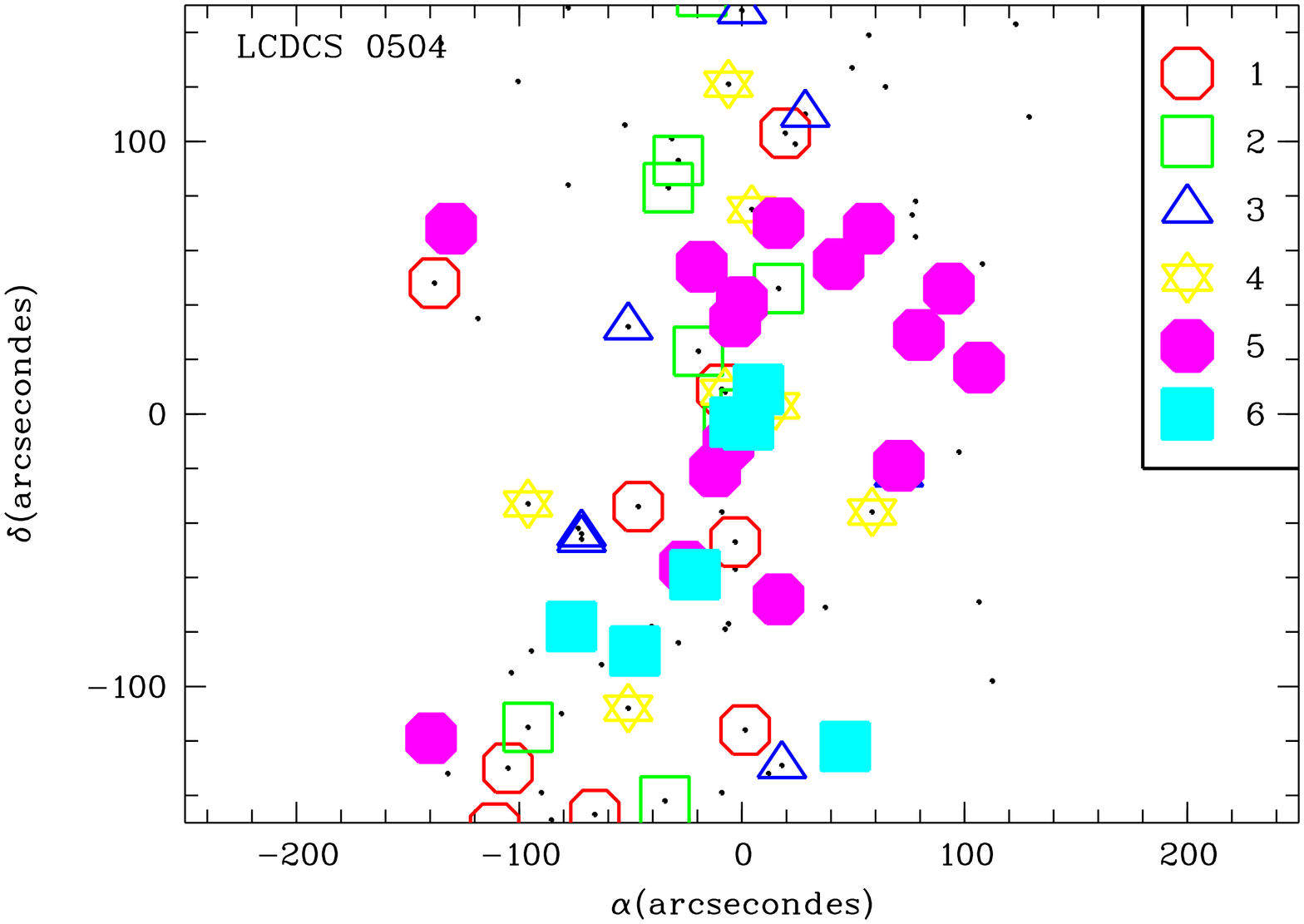}}
  }
  \centerline{
  \mbox{\includegraphics[angle=270,width=3.0in]{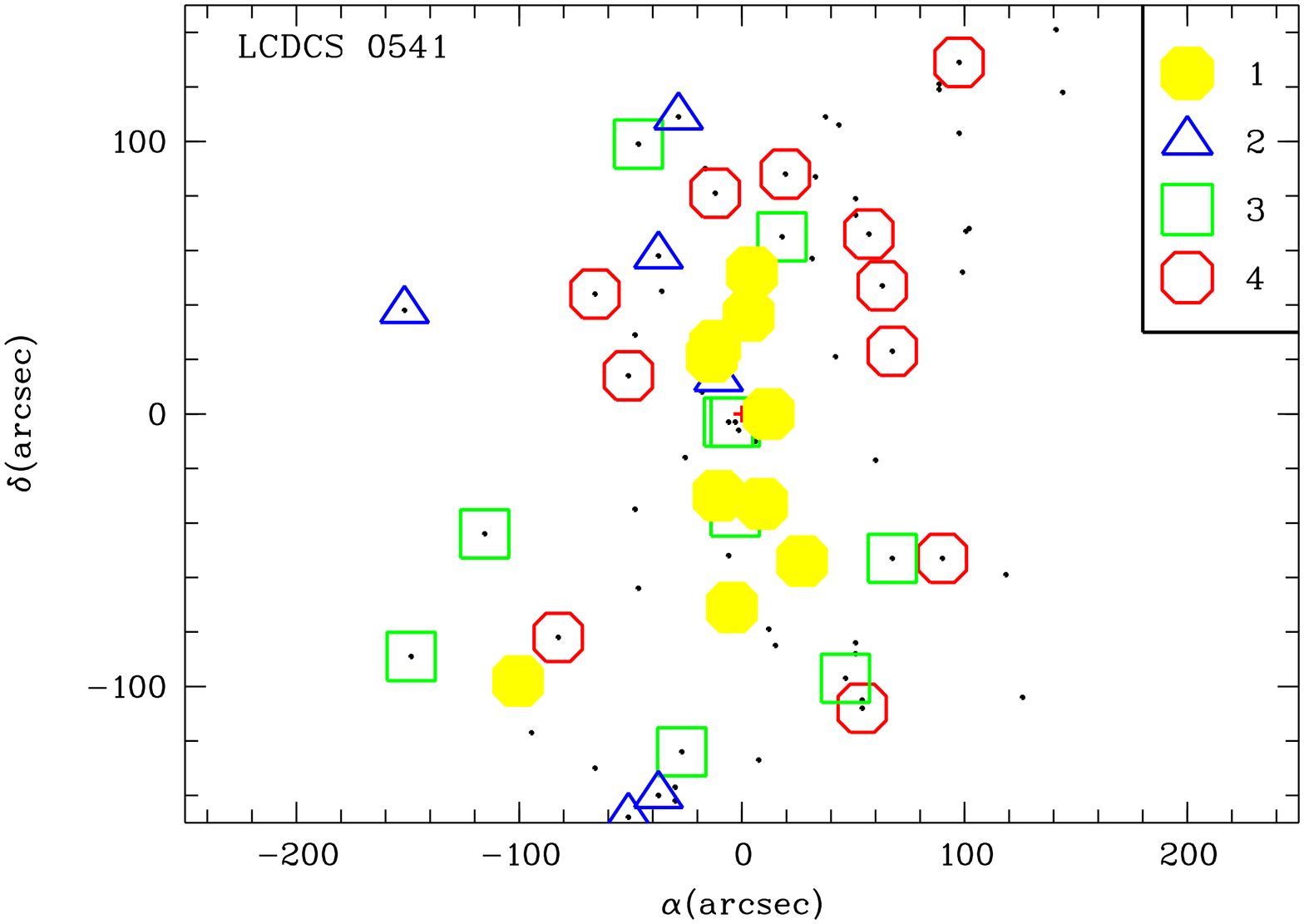}}
  \mbox{\includegraphics[angle=270,width=3.0in]{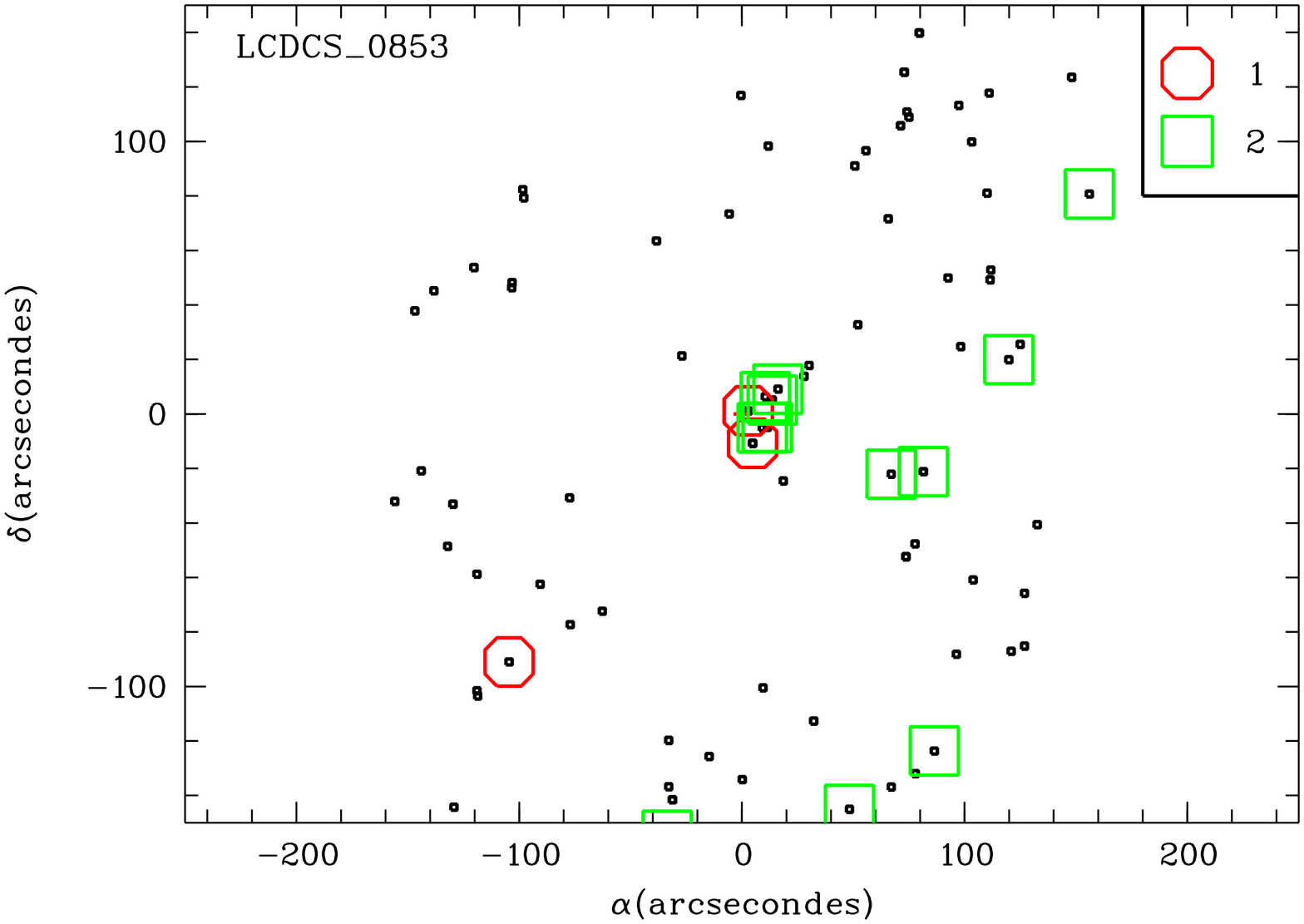}}
  }
  \caption{Maps of the detected cluster substructures with different
    symbols and colors representing the different structures. The symbol numbers refer
to each detected substructure.}
\label{ads}
\end{figure*}

We therefore analyzed the inner structure of the ten clusters using
the Serna-Gerbal method (Serna $\&$ Gerbal, 1996), which allows us to
both find and study substructures in galaxy clusters based on
dynamical arguments. This method adopts a hierarchical clustering
analysis to determine the relationship between galaxies based on their
relative binding energies (see Serna $\&$ Gerbal 1996 for a complete
description of the method). To determine whether a given galaxy is
part of a given cluster substructure or not, we need to know its
projected position, its magnitude, and its spectroscopic redshift (see
Table~\ref{sample} for the number of such available redshifts along
the considered lines of sight). We considered only substructures
containing at least three galaxies and within a redshift interval of
$\pm$0.012 from the cluster redshift (this is $\pm$3 times the maximal
velocity dispersion for a cluster). For two of the ten clusters, no
substructures were detected. This does not mean that these clusters
are fully relaxed: it is possible that our spectroscopic sampling was
insufficient to detect potential substructures.  For eight clusters,
significant substructures were detected and we show their spatial
distributions in Fig.~\ref{ads}.

The comparison of these subtructures
with the study of Halliday et al. (2004) (based on the
Dressler-Shectman test) is not straightforward.  However, limiting our
analysis to the clusters in common, we note that clusters with no
subtructures detected with the Serna $\&$ Gerbal method are found to
have no significant substructure in the Dressler-Shectman
test. Conversely, all clusters for which substructures are detected
with the Dressler-Shectman test also exhibit subtructures according to
the Serna $\&$ Gerbal method.

\begin{figure*}[!h]
  \centerline{
  \mbox{\includegraphics[angle=0,width=2.0in]{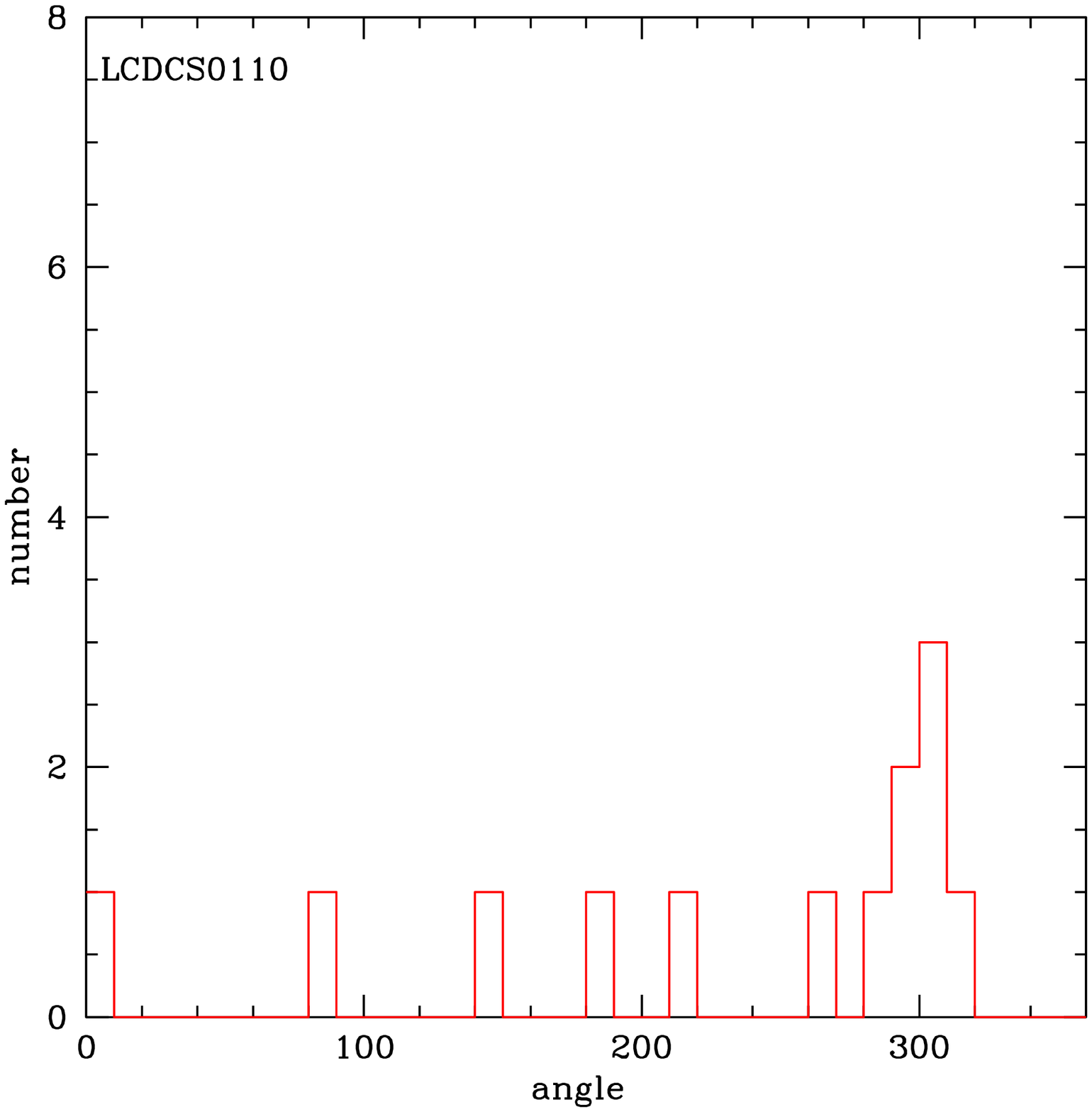}}
  \mbox{\includegraphics[angle=0,width=2.0in]{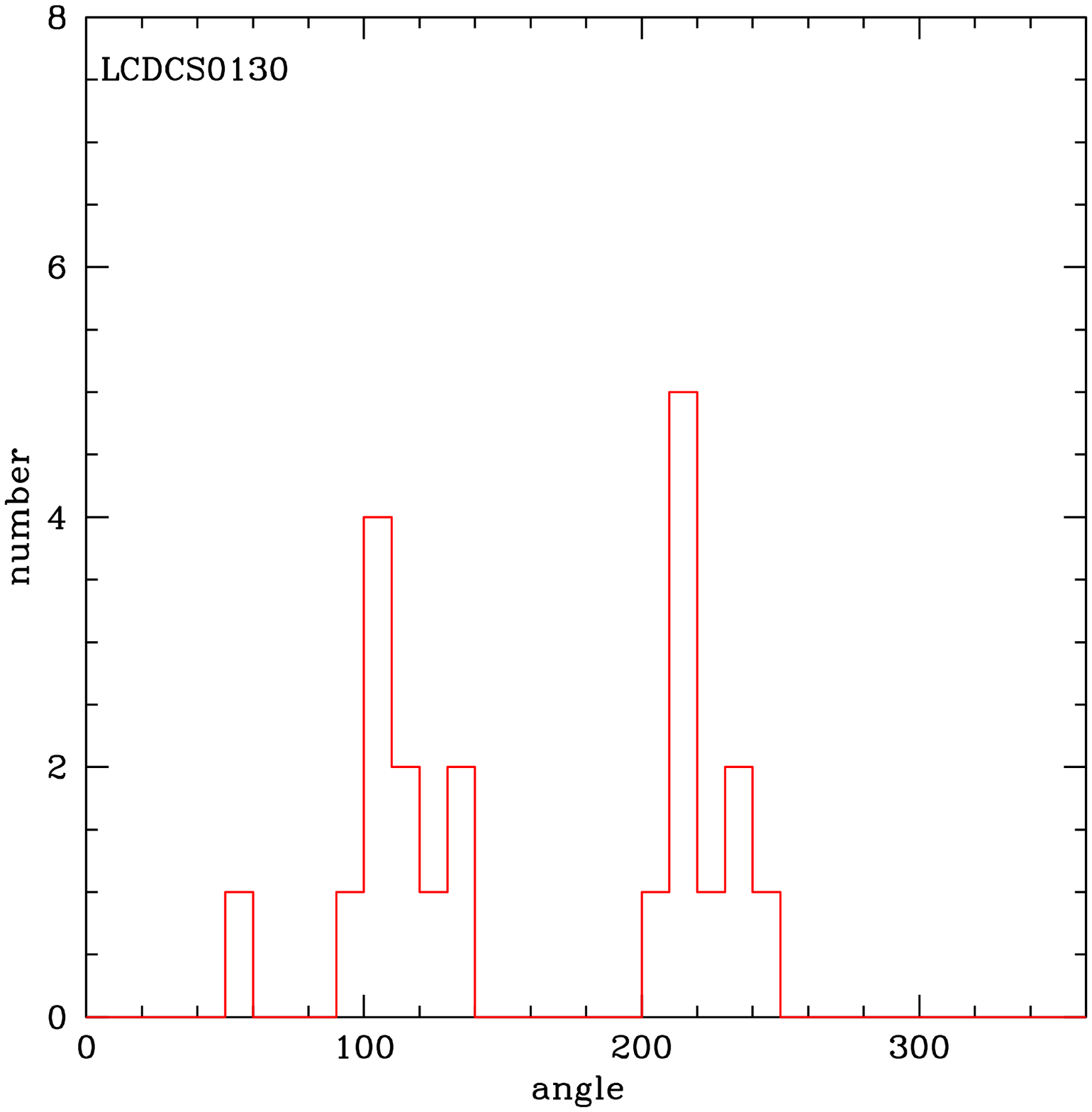}}
  }
  \centerline{
  \mbox{\includegraphics[angle=0,width=2.0in]{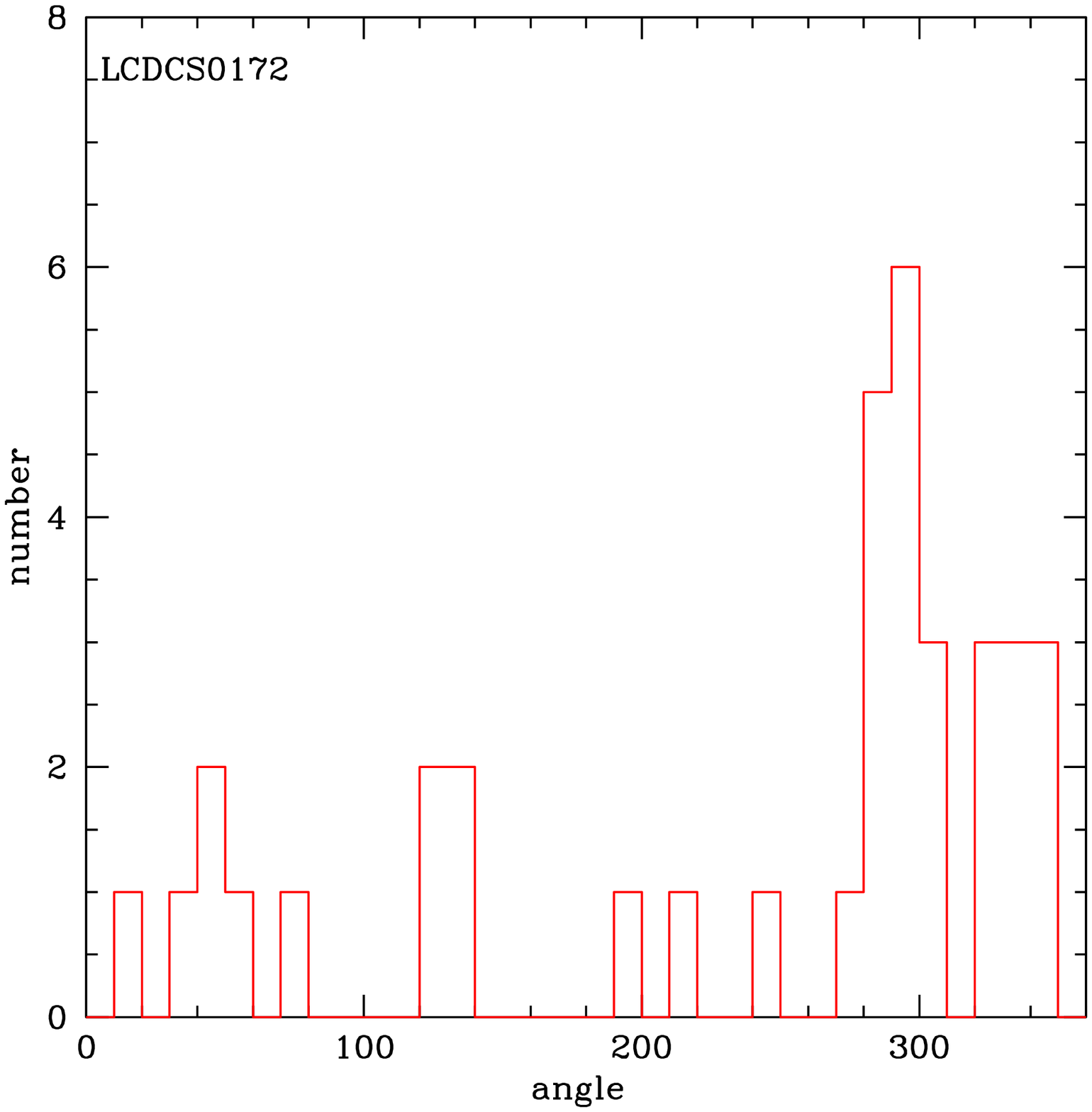}}
  \mbox{\includegraphics[angle=0,width=2.0in]{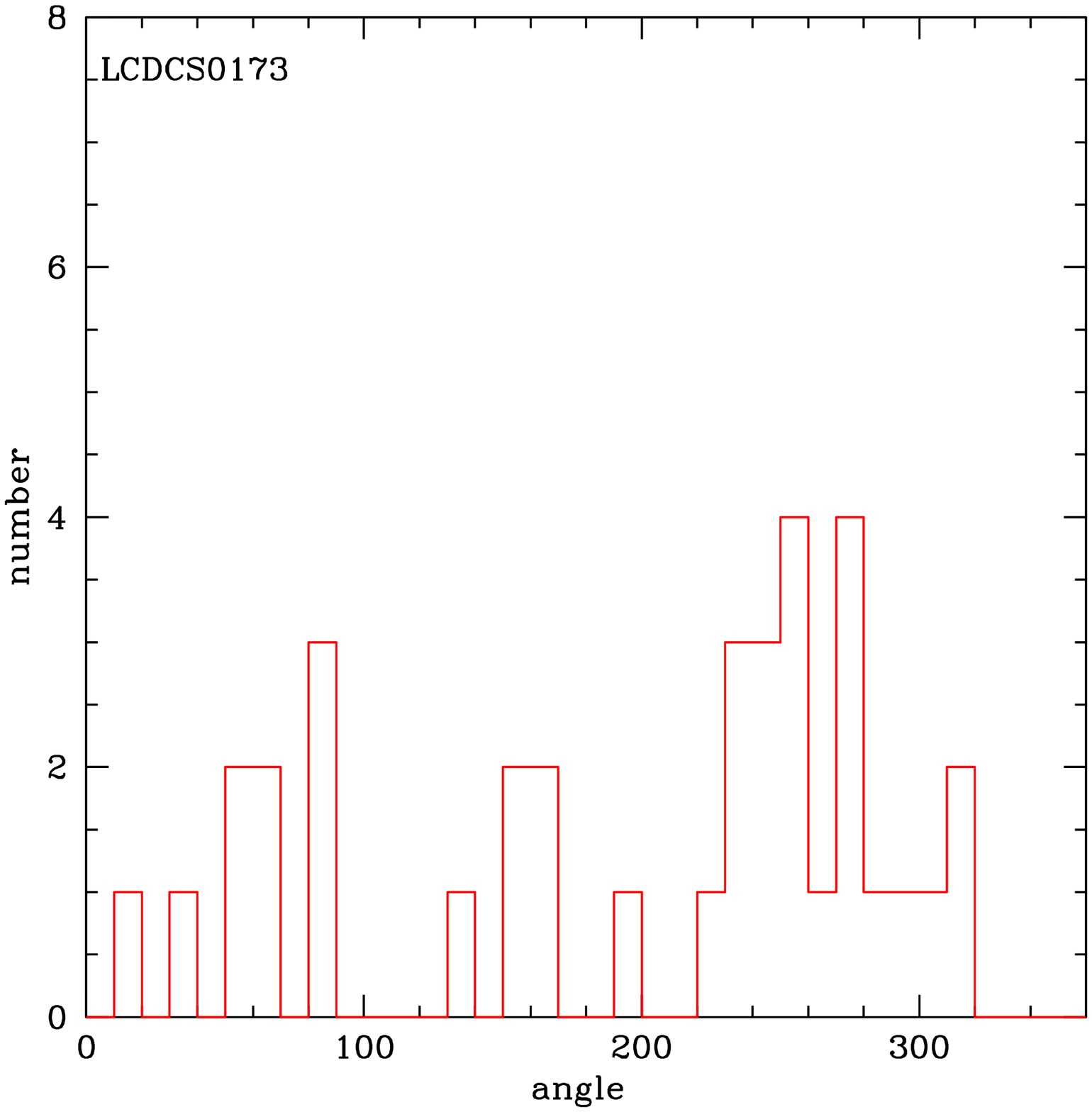}}
  }
  \centerline{
  \mbox{\includegraphics[angle=0,width=2.0in]{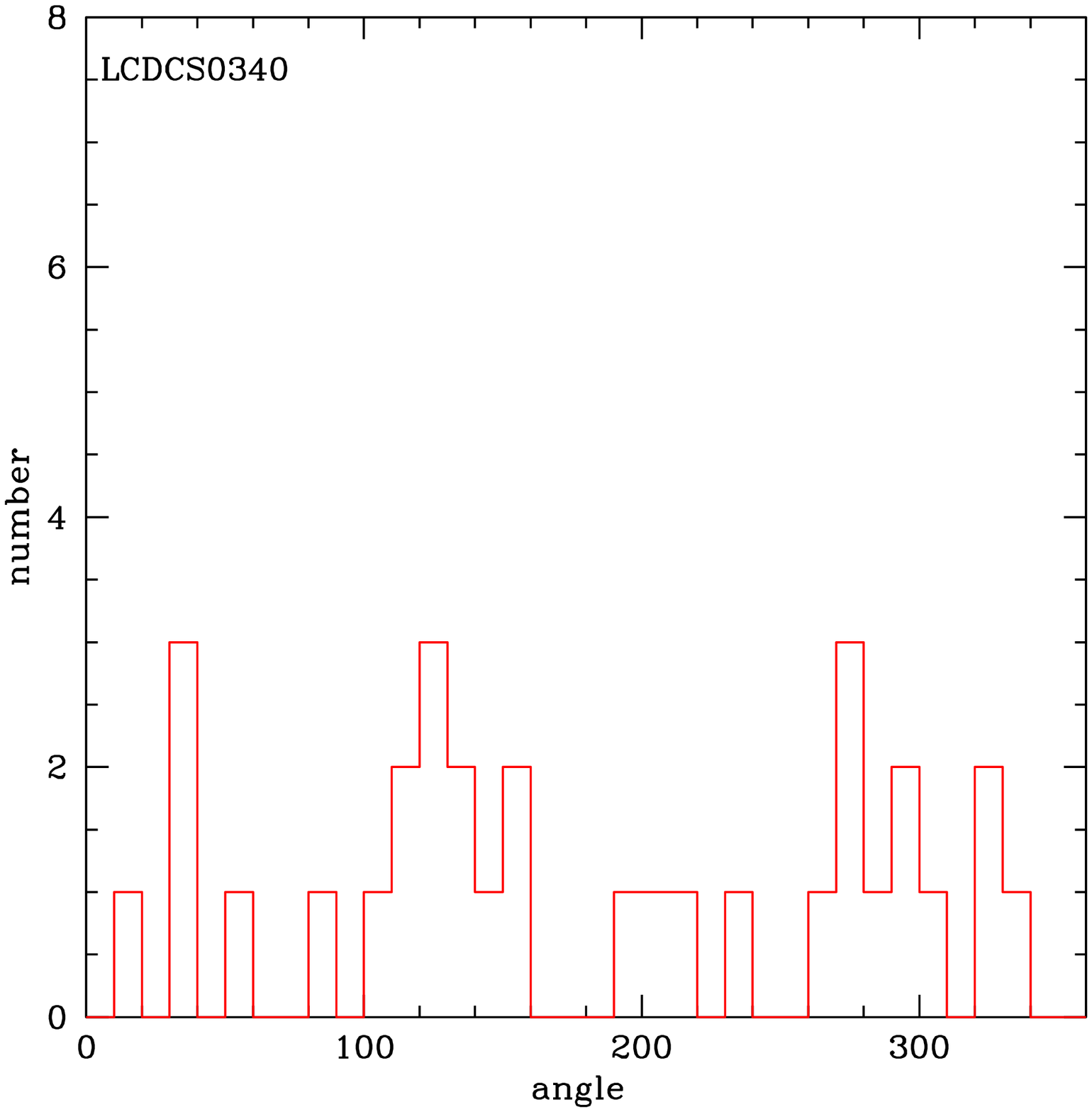}}
  \mbox{\includegraphics[angle=0,width=2.0in]{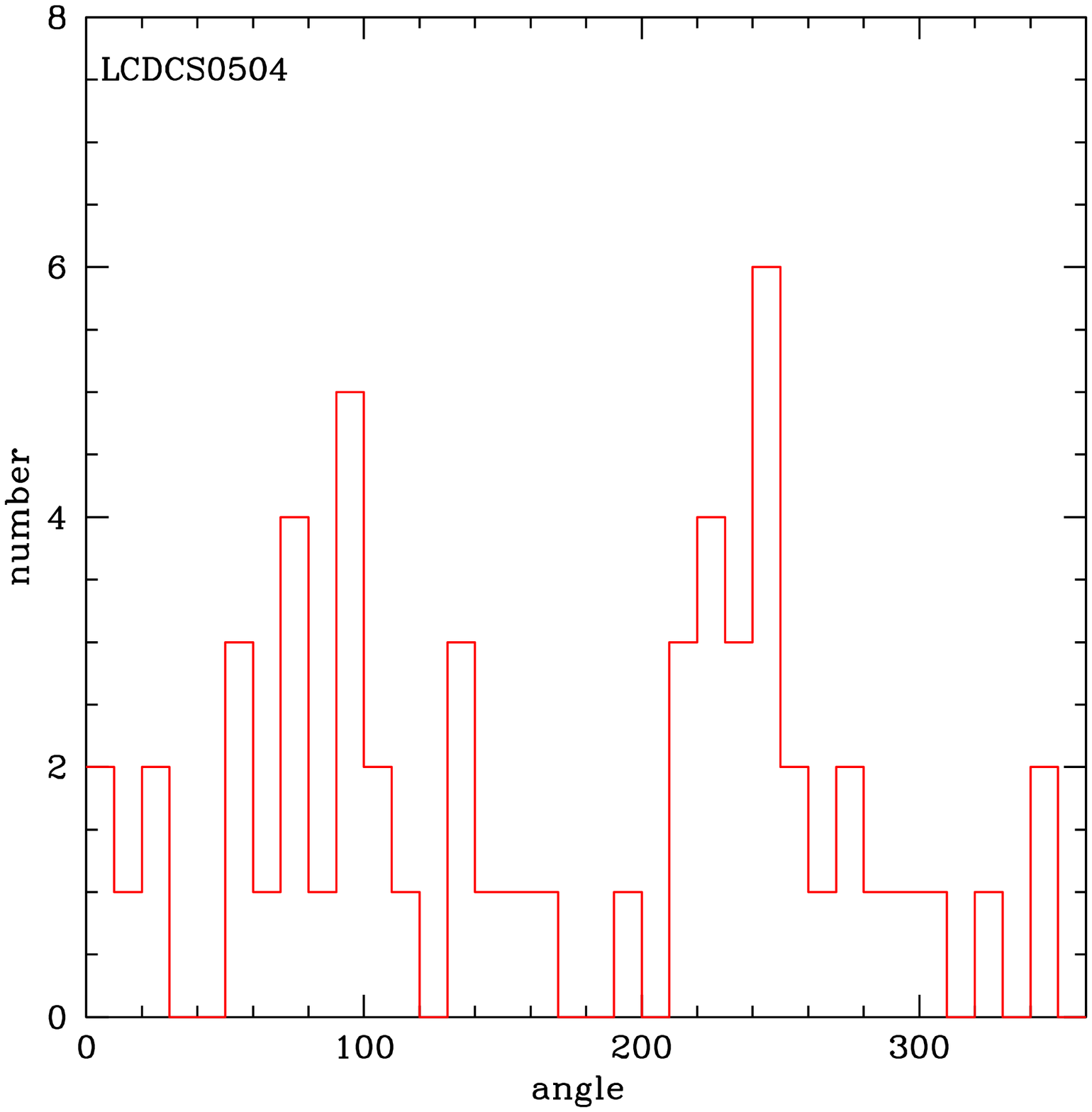}}
  }
  \centerline{
  \mbox{\includegraphics[angle=0,width=2.0in]{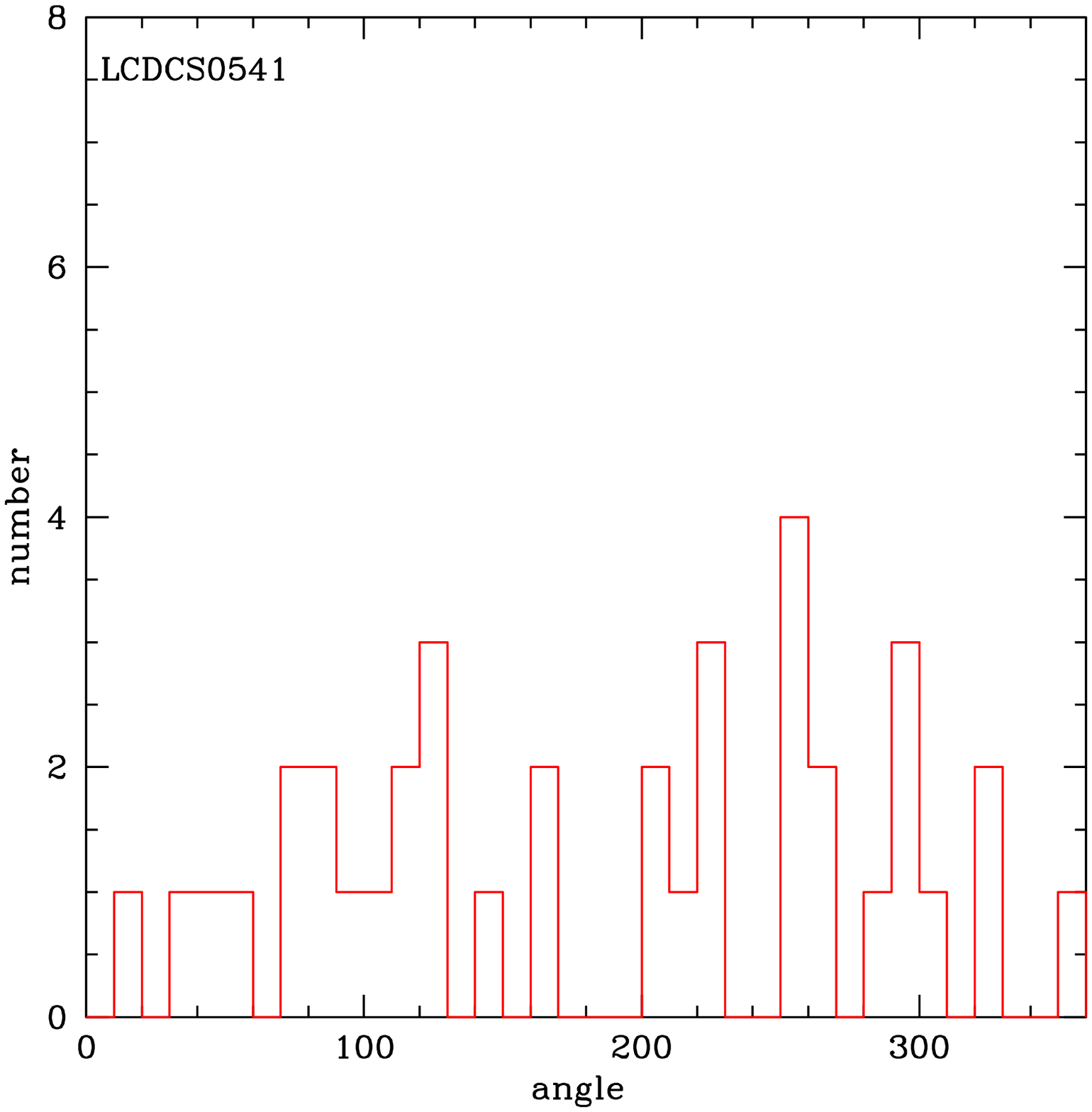}}
  \mbox{\includegraphics[angle=0,width=2.0in]{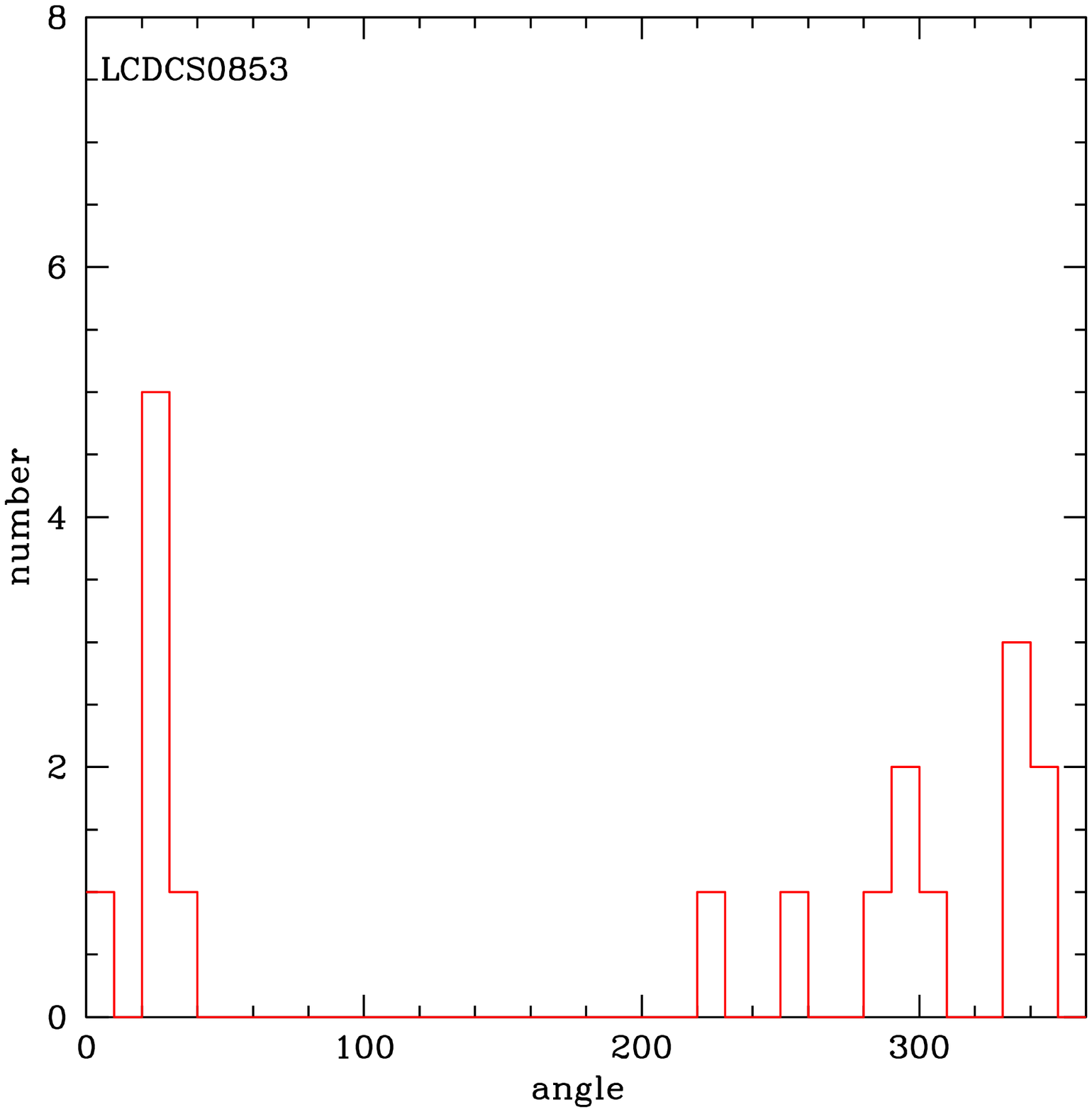}}
  }
  \caption{Histograms of the angular positions of the galaxies
    associated with substructures in the clusters
    analyzed. Preferential directions appear for LCDCS~0110,
    LCDCS~0130, LCDCS~0172, LCDCS~0504, and LCDCS~0853 (see text). Angles
are computed anticlockwise from West.}
\label{angle}
\end{figure*}

To help identify any cluster orientations, we computed histograms of
the angular positions of the galaxies found to be part of
substructures in the cluster (see Fig.~\ref{angle}).  We also tried to
infer the orientations directly from the ICL emission. However, these
sources are quite faint and any orientation estimate has a very large
uncertainty (of typically $\pm$70 deg).  Similarly, we could have
considered the orientation of the brightest cluster galaxy, but the
choice of this galaxy is not always obvious. We therefore decided to
rely only on the Serna-Gerbal results.  Clear preferential directions
appear for LCDCS~0110 (300~deg), LCDCS~0130 (220~deg), LCDCS~0172
(295~deg), LCDCS~0504 (240~deg), and LCDCS~0853 (25~deg).  The
  prefered orientation for LCDCS~0110 is mainly explained by
  substructure 1 (see Fig.~\ref{ads}). LCDCS~0130 shows a main
  orientation around 130 deg due to substructures 2 and 3, and there
  is a secondary orientation due to substructure 1 around 100~deg. The
  prefered orientations for LCDCS~0172 and LCDCS~0504 are due to all
  their substructures. Finally, the orientation of LCDCS~0853 is mainly
  driven by substructure 2.

Adopting the orientation directions determined by the Serna-Gerbal
method, we stacked the second $residual$ images computed with the
wavelet code using these same directions, and compared our results with
those for the randomly oriented stack. 
The ratio of these two image stacks does not vary by more than 5$\%$,
which means that, there is no significant detected orientation in the ICL
source distribution.

\subsection{Amount of ICL as a function of cluster velocity dispersion}

\begin{figure}[!h]
\begin{center}
\includegraphics[angle=270,width=3.0in]{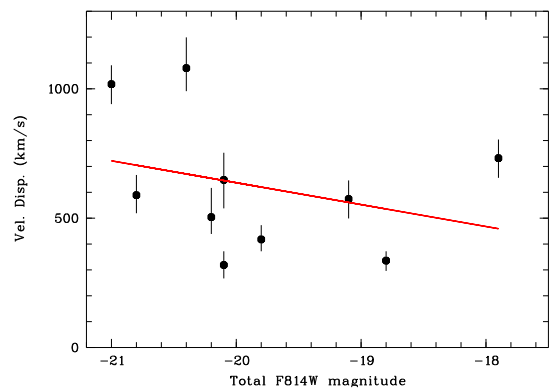}
\caption{Relation between the cluster velocity dispersions from
  Halliday et al. (2004) and Milvang-Jensen et al. (2008) and the
  amount of diffuse light in the F814W band. The red line has a slope
  of $-85$~km~s$^{-1}$~mag$^{-1}$. This figure does not show the low
  redshift clusters seen in Fig.~\ref{variz}.}
\label{varimass}
\end{center}
\end{figure}

\begin{figure}[!h]
\begin{center}
\includegraphics[angle=270,width=3.0in]{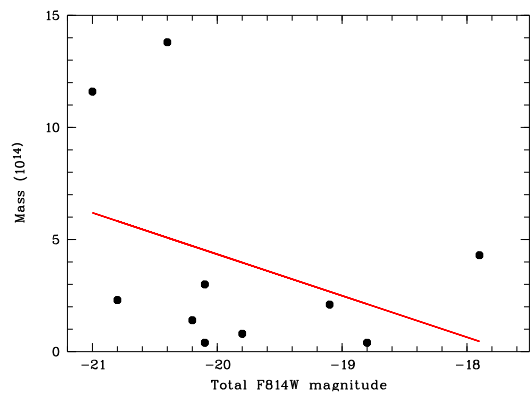}
\caption{Relation between the cluster $M_{200}$ mass computed from
  velocity dispersions from Table~\ref{sample} and the amount of
  diffuse light in the F814W-band. The red line has a slope of
  $-1.85$. This figure does not show the two low redshift clusters
  seen in Fig.~\ref{variz}.}
\label{varimass2}
\end{center}
\end{figure}

\begin{figure}[!h]
\begin{center}
\includegraphics[angle=270,width=3.0in]{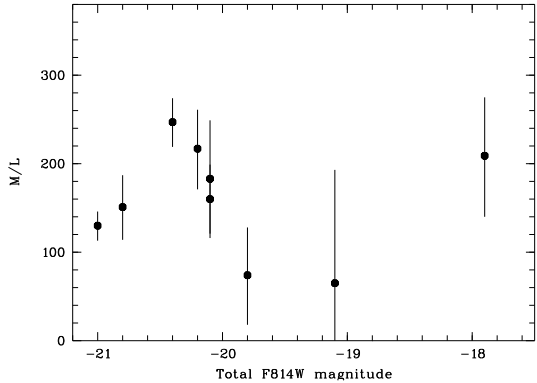}
\caption{Relation between the cluster M/L from Clowe et al.  (2006)
  and the total magnitude of the diffuse light in the F814W-band. This
  figure does not show the low redshift clusters seen in
  Fig.~\ref{variz}.}
\label{variML}
\end{center}
\end{figure}

An obvious question is whether the amount of diffuse light detected is
related to the cluster properties. We now consider ICL detections in
individual clusters. The  magnitudes quoted in Table~1 are obviously
lower limits, only characteristic of the diffuse light we are able to
detect given our instrumental configuration. We show in
Fig.~\ref{varimass} the relation between the cluster velocity
dispersions (from Halliday et al. 2004 and Milvang-Jensen et
al. 2008) and the amount of diffuse light (see Table~\ref{sample}).
We note here that these velocity dispersions are reasonably close to
the Clowe et al. (2006) estimates. We see a modest increase in
  the F814W absolute magnitude of the diffuse light with the velocity
  dispersion, mainly because of the two highest velocity dispersion
clusters in the considered sample. In Fig.~\ref{varimass2}, we
similarly show the relation between the amount of diffuse light and
the cluster mass. We estimated the mass (see Table~\ref{sample}) using
the velocity dispersions and Eq. 6 of Evrard et al. (2008). We still
see a modest increase in the total F814W absolute magnitude of the
diffuse light with mass, mainly because of the two highest mass
clusters.  We attempted to identify any obvious relation between the
cluster M/L and the amount of ICL. We were unable to find any clear
trend (see Fig.~\ref{variML}) and simply note that the amount of dark
matter in a cluster does not appear to strongly influence the amount
of material expelled from galaxies that probably forms the ICL (see
also below). This is in good agreement with the simulations of Dolag
et al.  (2010), which do not predict any significant relation between
the cluster mass and the fraction of diffuse light in clusters.

\subsection{Comparison with low redshift detections}

We now place our results in perspective with our previous
studies of less distant clusters, namely Coma and A2667 (see Adami et
al. 2005 and Covone et al. 2006a respectively). The detection of
diffuse light in these two structures has been made with the same wavelet
method and the results are therefore directly comparable.  In Coma
(z=0.023), Adami et al. detected diffuse light sources equivalent to a $-22.6$
galaxy in the F814W filter. This value was estimated by translating
the Adami et al. (2005) R band Vega value to an F814W AB value using
the Fukugita et al. (1995) k-corrections (0.05 mag at the Coma cluster
redshift) and elliptical galaxy colors. In A2667 (z=0.233), we
detected diffuse light sources equivalent to a $-21.12$ galaxy in the
F814W filter (adopted k-correction: 0.25 mag).

These values allow us to search for any evolution in the amount of ICL
in clusters between z$\sim$0.8 and z=0. Before trying to plot the
amount of ICL in clusters as a function of redshift, we must correct
for the dependence on velocity dispersion and mass discernible in
Figs.~\ref{varimass} and ~\ref{varimass2}. The best-fit relation
between the velocity dispersion $\sigma _v$ (respectively the mass)
and the absolute magnitude of the diffuse light in the F814W filter is
$$ \sigma _v = -85 M_{F814W} - 1056 $$ 
and
$$ Mass_{200} = -1.85 M_{F814W} - 32.66 $$ 

\begin{figure}[!h]
\begin{center}
\includegraphics[angle=270,width=3.0in]{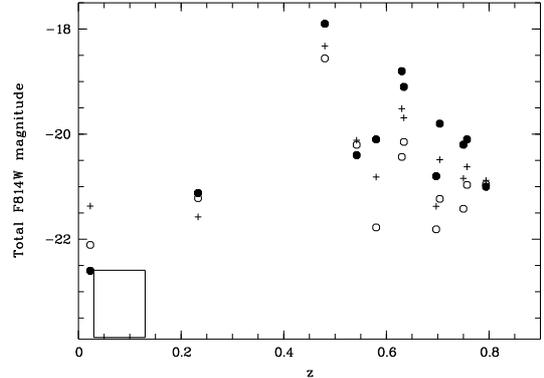}
\caption{Relation between the total magnitude of the diffuse light in
  the F814W band and the cluster redshift. Filled symbols are raw
  values. Open circles are the ICL total F814W magnitudes corrected for
  the effects of velocity dispersion.  Crosses are the values
  corrected for the effects of mass.  The large square at the bottom
  left of the figure represents the locus occupied by intracluster
  light detected in the Gonz\'alez et al. (2005) low-redshift clusters
  (see text).}
\label{variz}
\end{center}
\end{figure}

If we assume that the same relation applies to Coma and A2667
(adopting velocity dispersions of 1200 km/s for Coma from Adami et al. 2009
and 960 km/s for A2667 from Covone et al. 2006b), we can correct
for the differences in velocity dispersion between all the clusters
considered here. Adopting the cluster masses of Kubo et al. (2007) and
Ota $\&$ Mitsuda (2004) for Coma and A2667, we can also correct
for any dependence on mass. We show in Fig.~\ref{variz} the variation of the
total (rest-frame) ICL magnitude versus redshift. We see no strong
variation in the amount of ICL for the clusters between z=0 and z=0.8,
especially when considering the mass-corrected values, besides perhaps a 
modest increase with redshift.

Several other studies have been performed at low redshift, some of which 
based on galaxy subtraction via their shape modeling (e.g. Gonz\'alez et al. 2007,
Krick \& Bernstein 2007, Rudick et al. 2010). These studies sometimes predict more
diffuse light in structures than the studies based on the wavelet method (e.g.
Adami et al. 2005, Covone et al. 2006a, Da Rocha et al. 2008). 

For example, we can use the results of Gonz\'alez et al. (2005) to make
this comparison. By fitting several profiles of the brightest galaxies
of 24 z=[0.03,0.13] galaxy clusters, these authors estimated the
total magnitude included within several radii. This definition differs
from the one adopted here (we do not use, for example, any parametric
model of the diffuse light). However, we can adapt these values by
removing the amount of light included in the 10~kpc radius from the
value included in the 50 kpc radius of Gonz\'alez et al. (2005) (see
their Table 6). Assuming a k-correction factor of 0.07 (Gonz\'alez et
al. 2005), a F814W-I of 0.04 (Fukugita et al. 1995), and an AB-Vega
correction factor of 0.46, we then plot in Fig.~\ref{variz} the locus
occupied by the Gonz\'alez et al. (2005) clusters. The diffuse light
values are somewhat brighter than our present detections, perhaps
because our typical ICL sources radii are close to 20~kpc, hence less
extended than the 50~kpc of Gonz\'alez et al. (2005).

\section{Nature of the wavelet-detected sources}

\subsection{ICL sources}

An important question is whether the diffuse light that we have detected
consists of stars or diffuse gas. A common assumption is that we are dealing
with old stars expelled from disrupted or partially disrupted galaxies. For
example, Adami et al. (2005) discuss this and show that the diffuse light in
Coma has colors that closely agree with those of quite old stellar populations.
However, can this diffuse light be produced by hot intra-cluster diffuse gas?

The bremsstrahlung emission of the hot gas commonly detected in X-rays
might also be detectable at optical wavelengths. It is therefore
interesting to compare the surface brightness we have observed for the
ICL with the predicted value for hot gas in a cluster with an X-ray
luminosity similar to that of our sample.  We selected the Coma
cluster and carried out an order of magnitude estimate based on
Lutovinov et al. (2008). Within a 10~arcmin radius, they found about
half the total X-ray luminosity (total flux =$4.4 \times 10^{-10}$
ergs cm$^{-2}$ s$^{-1}$). We then used the PIMMs code (see
http://heasarc.nasa.gov/docs/software/tools/pimms.html) to convert the
50\% total luminosity to 2.5-3 eV (approximately B band) and a
10~arcmin radius to derive a surface brightness of about $1.13\
10^{-16}$ erg cm$^{-2}$ s$^{-1}$ for an area of 300~arcmin$^2$. We
then converted this to AB B magnitudes using IDL to obtain a value of
about 34 mag (AB)/arcsec$^{-2}$ (typical values for our ICL sources
are $\sim$27 mag/arcsec$^2$).  We also found similar values for the V
and I bands, much fainter than the brightnesses we can detect. Therefore,
the optical tail of the thermal bremsstrahlung responsible for the
cluster X-ray emission cannot account for the ICL emission that we
measure.

Hence, the most likely interpretation is that we observe stars expelled from
cluster galaxies located at the bottom of the cluster potential well.

Additional arguments for the ICL consisting of stars are that the
intracluster supernova rates are consistent with the size of this
population of stars (Sand et al. 2011), the baryon budget being close
to the WMAP universal value if one adopts a standard stellar M/L for
this light (including stars in galaxies and hot plasma; cf. Gonz\'alez
et al. 2007), and that stars (PNe and red giants) have been detected
in-between galaxies in clusters (cf. Arnaboldi et al. 2010). We also
note that our V-band images are deep enough to detect WDCOs (see next
section) for not too distant clusters, but are insufficiently deep and with 
an unadapted background subtraction to detect ICL sources. We therefore do 
not have access to the colors of the ICL sources detected in our F814W images.

\subsection{Wavelet-detected compact objects (WDCOs)}

During its first two passes, the wavelet detection process appears to
detect very faint objects in the HST ACS F814W images, that are
sometimes not detected by classical SExtractor searches (unless we use
very low thresholds (as low as 0.3), which would lead to numerous
fake detections). These faint objects could be fake detections,
and even if real cannot be characterised reliably using data in a
single band. To check whether these objects are real and investigate
their nature, we analyzed the three lowest redshift clusters in our
LCDCS sample for which our ground-based V-band images are deep
enough. We then selected only WDCOs detected in both the V and F814W
images. These WDCOs are most probably real and not caused by random
background fluctuations.

\begin{figure}[]
\mbox{\includegraphics[angle=0,width=2.3in]{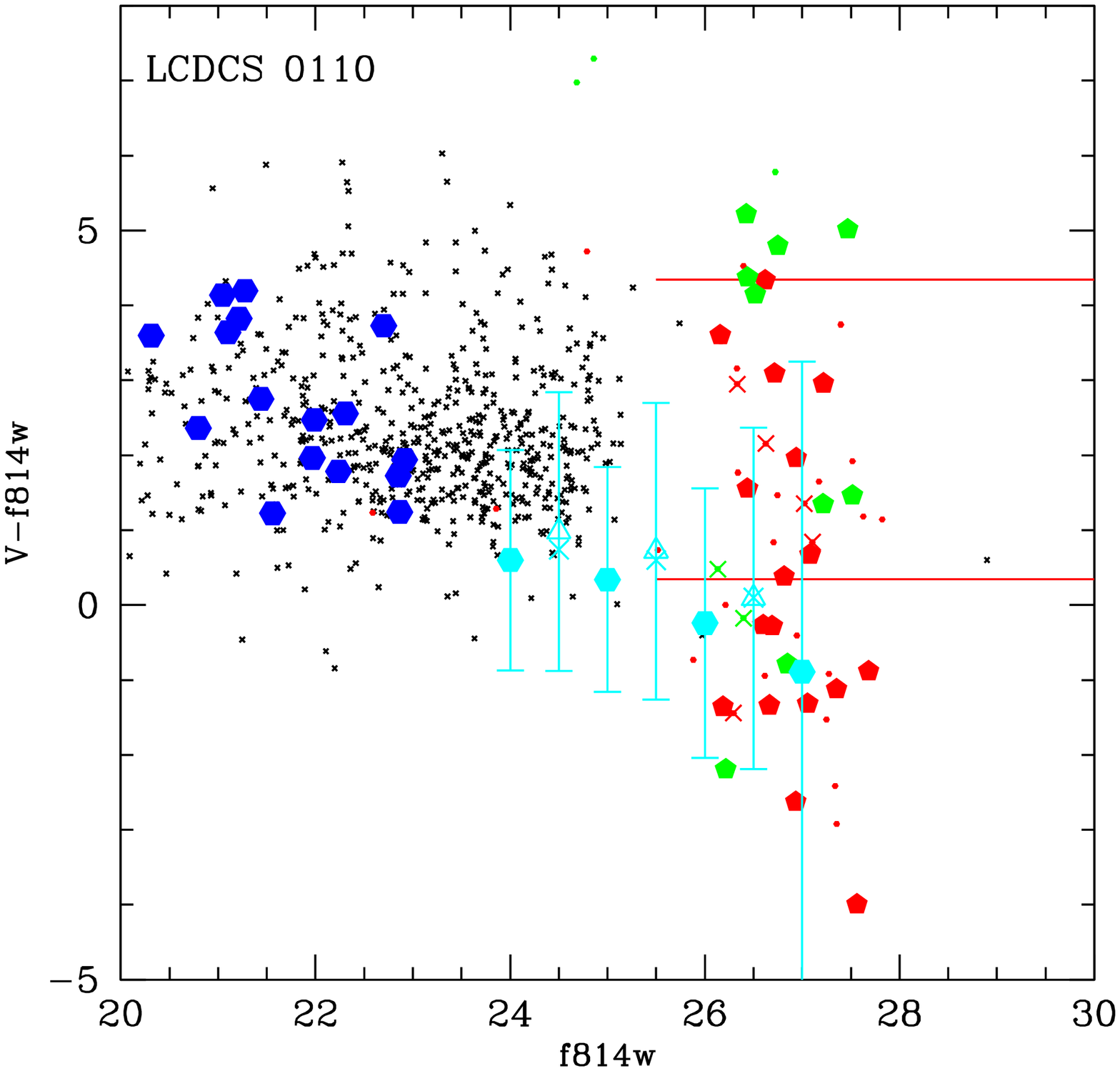}}
\mbox{\includegraphics[angle=0,width=2.3in]{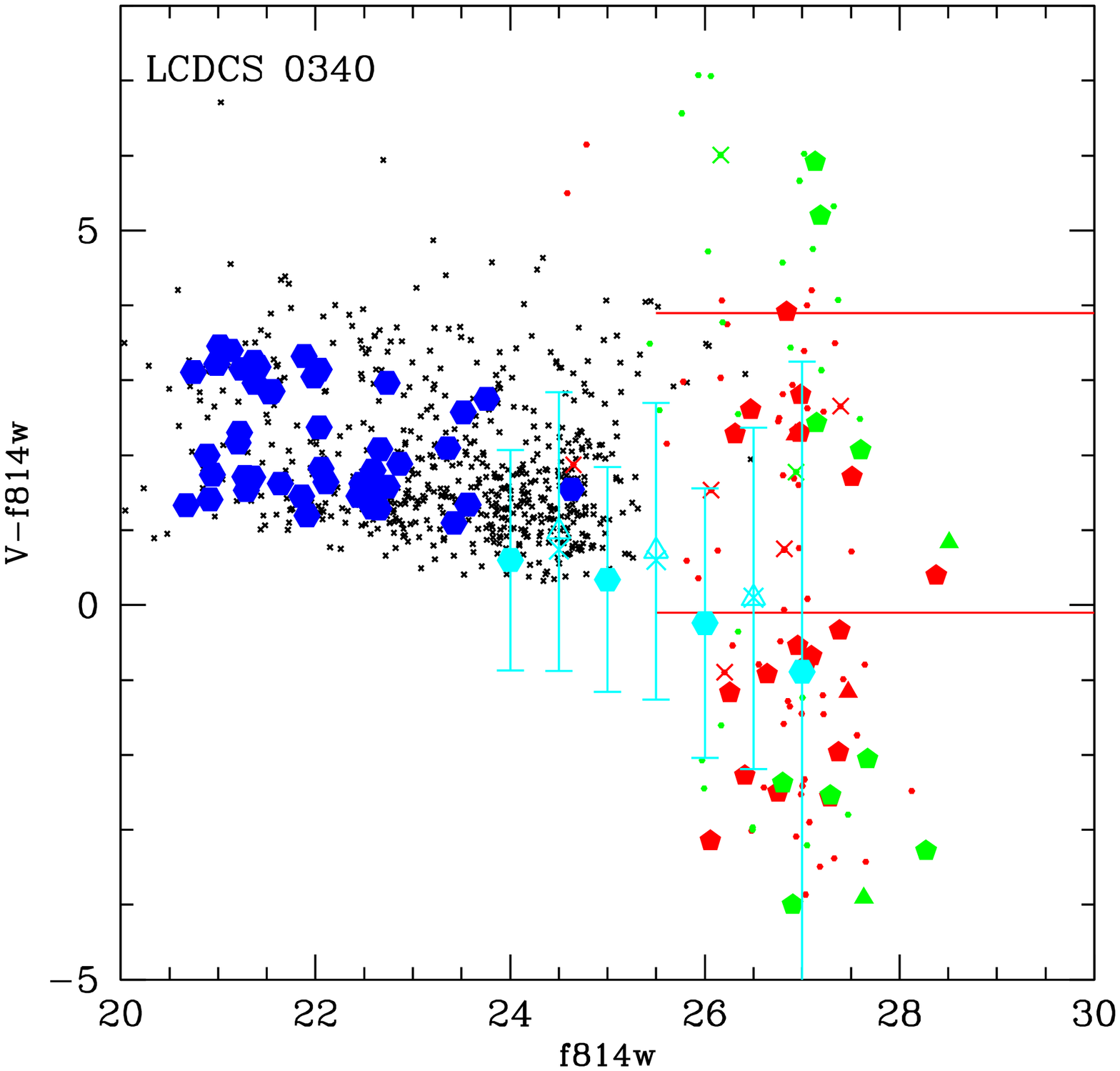}}
\mbox{\includegraphics[angle=0,width=2.3in]{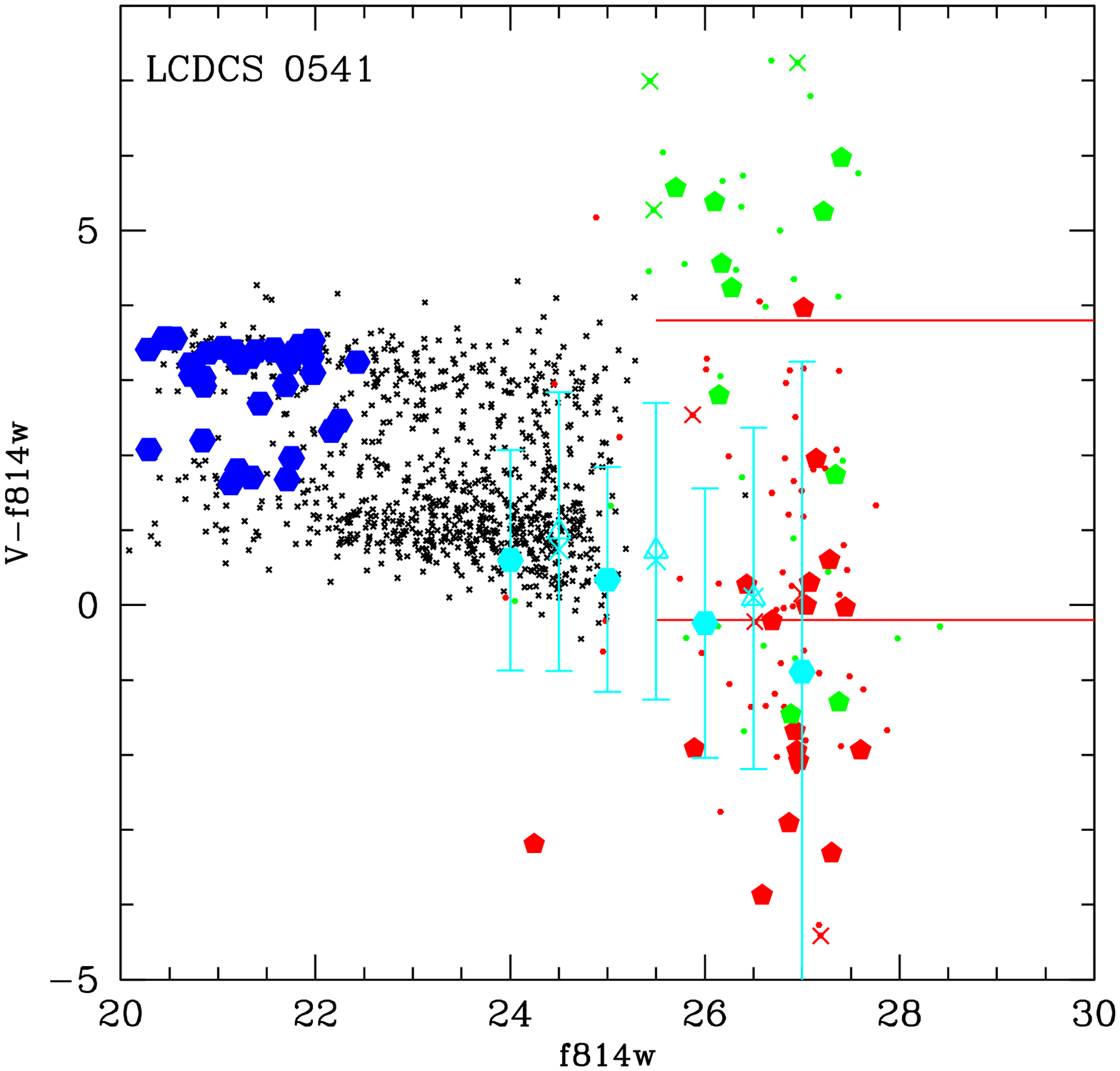}}
\caption{Color-magnitude diagram V-F814W versus F814W for the three
  considered clusters. Black dots represent the SExtractor-detected
  objects along the considered lines of sight. Blue-filled hexagons
  are the SExtractor-detected objects in each cluster according to
  their spectroscopic redshift. Triangles are WDCOs with a $\chi ^2$
  lower than 8, crosses with a $\chi ^2$ between 8 and 10, pentagons
  with a $\chi ^2$ between 10 and 15, and small dots with higher $\chi
  ^2$ values. For these last four classes, red is the color for WDCOs
  detected in the first pass and green in the second pass. The red
  lines limit the space in which the objects are likely to belong to
  the cluster. The large cyan filled circles with error bars are the
  locus of galaxies in the VIRMOS 2hours field (see text). The values
  of these four cyan circles were computed (from left to right) with
  29689, 35804, 4788, and 511 galaxies.  The large cyan open triangles
  with error bars are for the galaxies in the VIRMOS field with mean
  surface brightness typical of the first pass WDCOs.  The large cyan
  crosses without error bars are for the galaxies in the VIRMOS field
  with mean surface brightness typical of the second pass WDCOs.}
\label{fig:diagram}
\end{figure}

To see if these objects are peculiarly distributed in color-magnitude
plots (see Fig.~\ref{fig:diagram}), we have to compare them with deep
sky catalogs. These catalogs must include V and F814W-like magnitudes
in order not to be biased by different bands. thus excluding for
example deep fields observed with SDSS filters. They also must be deep
enough to be comparable with our own V and F814W data. We chose to use
the VIRMOS deep imaging survey 2hours field (see e.g. McCracken et
al. 2003). This field was observed in the B, V (very close to our own
V band), R, and I bands at the CFHT telescope with the CFH12K
camera. Our F814W band can easily be mimicked by the I band at the
cost of a $\sim$0.1 magnitude shift, nearly constant as a function of
redshift and galaxy type (Fukugita et al.  1995). Moreover, the VIRMOS
deep imaging survey 2hours field is one of the very rare fields deep
enough to be comparable with our data. Fig.9 of McCracken et
al. (2003) shows that the detection rate is still non negligible at
I$\sim$26.5/27, even though beyond the 90$\%$ completeness magnitude. We
therefore plot in Fig.~\ref{fig:diagram} the mean V-F814W color of the
VIRMOS deep imaging survey 2hours field as well as its $3\sigma$
uncertainty, between F814W=24 and 27. The involved color ranges are in
perfect agreement with most of the WDCOs.

However, the objects in the VIRMOS deep images may occupy a
  different locus in a mean surface brightness versus total magnitude
  plane compared to our WDCOs. We estimated the mean surface
  brightnesses of our first and second pass WDCOs in the same way as
  the surface brightnesses computed for the VIRMOS deep imaging survey
  2~hour field. This was done by setting a low detection threshold in
  SExtractor and by keeping only the detections lying on the 
  WDCOs positions. We then computed the mean surface brightnesses of
  these objects with the parameters estimated by SExtractor in an area
  defined by 3 times the minor and major axes. First (resp. second)
  pass WDCOs then have SExtractor-estimated mean surface brightnesses
  of 25.2$\pm$0.53 (resp. 25.6$\pm$0.45). We applied these two surface
  brightness selections to the VIRMOS deep imaging survey 2~hour field
  catalog before recomputing the V-F814W colors.
  Fig.~\ref{fig:diagram} shows that WDCOs still appear not to have
  atypical colors.

However, the VIRMOS deep imaging survey 2hours field does not include
very massive clusters (see e.g. Adami et al. 2011). In order to
compare the WDCO colors with very faint galaxies of massive clusters,
we defined an approximate locus for the red-sequence in the WDCO
regime assuming the mean V-F814W color from faint cluster member
galaxies defined in a deep Coma cluster spectroscopic redshift catalog
(see Adami et al. 2009). We assumed a 1$\sigma$ red envelope width of
0.7 at magnitudes between F814W=26 and 28 (leading to a 3$\sigma$
width of $\pm$2.1), based on the typical cluster red-sequence
1$\sigma$ width for Coma at R=22 being close to 0.5 (see Adami et
al. 2009). Given the lack of knowledge of the red sequence width at
F814W$\geq$26, we adopted the simplest approach of considering a
horizontal envelope for the red sequence. We see in
Fig.~\ref{fig:diagram} that part of the WDCOs are also located on the
red-sequence defined above. The WDCOs with colors consistent with the
cluster red-sequence could therefore be small, very faint, compact
galaxies and members of the considered clusters (at least part of
them).

\begin{figure}[h!]
\begin{center}
\includegraphics[angle=0,width=3.00in]{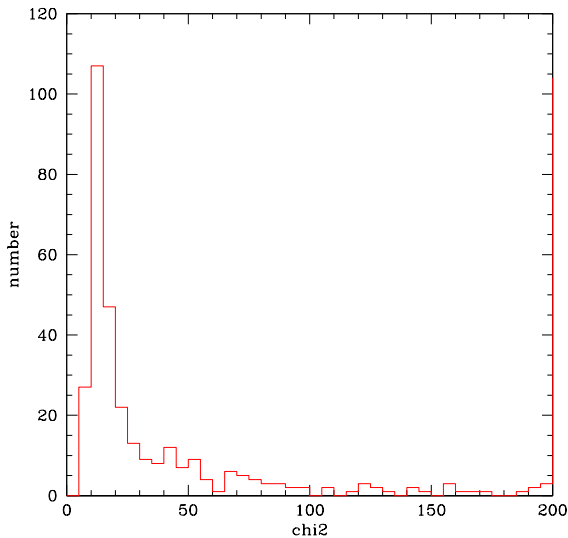}
\caption{Histogram of the reduced $\chi ^2$ values obtained by
    merging the results for the three clusters.}
\label{fig:histosep}
\end{center}
\end{figure}

We also note that most of the WDCOs are more than one magnitude
fainter than the SExtractor detection limit, and remarkably, nearly
none of them are brighter than this limit. Therefore, they appear to
be an independent population in the three considered clusters or might
also be atypically faint Galactic stars. We cannot use SExtractor to
make a classification since these WDCOs are by definition not detected
by this code. We are similarly unable to place the WDCOs in a central
magnitude versus total magnitude plot because WDCOs usually have very
irregular brightness profiles, hence a reliable computation of the
central brightness is impossible. We therefore fitted a
two-dimensional Gaussian with an added constant background to each of
the WDCO images. We assumed that the full-width at half maximum (FWHM)
of the Gaussian was the HST ACS spatial resolution and only allowed
the level of the background and the amplitude of the Gaussian to
vary. We show in Fig.~\ref{fig:histosep} the histogram of the reduced
$\chi ^2$  computed for the best fits.

All objects have quite large reduced $\chi ^2$ values, making them
very unlikely to be faint stars. Even the lowest reduced $\chi ^2$
values are on the order of 8, much too high for this population
to be dominated by objects limited by the instrumental resolution.

\begin{figure}[h!]
\begin{center}
\includegraphics[angle=270,width=3.00in]{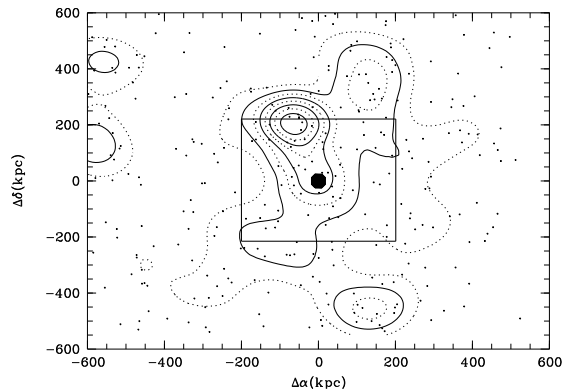}
\caption{Upper figure: stacked physical spatial map of the WDCOs
  detected in the clusters analyzed in both V and F814W bands. Density
  contours start at the 3$\sigma$ level and are in steps of
  1$\sigma$. The area limited by the inner square is shown in the
  lower figure. Lower figure: stacked and rescaled images of
  LCDCS~0110, LCDCS~0340, and LCDCS~0541 (see text). Red circles are
  the detected WDCOs. The two blue lines delimit a lower signal to
  noise region.}
\label{fig:kernel}
\end{center}
\end{figure}

We now consider the redshifts of these WDCOs: are they part of the
considered clusters? It is obviously impossible to measure their
spectroscopic redshifts with the telescopes presently
available. Photometric redshifts are also impossible to compute
because these objects are only detected in two bands and our
photometry in the other optical bands is insufficiently deep.  Beyond
the color-magnitude relations already discussed, another way to
investigate the cluster membership of these objects is then to study
their spatial distribution.  If cluster members for part of them,
WDCOs should at least partially follow the light of the cluster
potential well.  Fig.~\ref{fig:kernel} clearly shows such a
behavior. To generate this figure we used an adaptive kernel technique
(e.g. Adami et al. 1998) and computed the projected density map on the
sky of all the detected WDCOs along the line of sight. We selected a
structure-free area and computed the 1$\sigma$ variation in the WDCO
density. Dividing the original WDCO density map by this value, this
produced Fig.~\ref{fig:kernel}, which is expressed in units of this
1$\sigma$ variation.  We clearly see a central concentration with a
scale typical of a cluster (we used the mean redshift of the
considered clusters to give Fig.~\ref{fig:kernel} in units of
kpc). If WDCOs are real, this is tempting evidence in favor of
  these objects being cluster members.

  Despite the fact that the considered WDCOs are detected in two
  different bands observed with two different instruments, it could be
  argued that since they are such low signal to noise objects they
  could well be purely noise at least partly occuring at the edges of
  bright objects and/or in noisy regions. We would then naturally
  expect these ``spurious'' detections to be located near the haloes
  of bright objects. They would then follow the cluster profile and
  this could explain why their spatial distribution is centrally
  concentrated. However, we first note that the spatial distribution
  of WDCOs is not exactly centered on the cluster centers: a
  $\sim$200~kpc shift is seen in Fig.~\ref{fig:kernel}. Second,
  Fig.~\ref{fig:kernel} also shows the location of the WDCOs in the
  stacked central cluster regions. We made the sum of the images of
  LCDCS~0110, LCDCS~0340, and LCDCS~0541 after centering each image on
  the cluster center and rescaling to physical units (kpc).  While a
  few of the WDCOs are indeed located in the vicinity of bright
  objects and may well be spurious objects, the vast majority of WDCOs
  appears to be far from any bright object. Moreover, the higher noise
  part of images shown in Fig.~\ref{fig:kernel} exhibits only two
  WDCOs. It is therefore very likely that the large majority of our
  WDCOs is not made of spurious detections.

  We made a last test: in the external parts of our deep HST images of
  LCDCS~0110, we considered a region populated with bright objects
  (including stars), but without any known cluster. If WDCOs were
  spurious objects wrongly detected at the edges of bright objects, we
  should then detect significant concentrations of WDCOs in these
  areas around bright objects, despite the absence of any
  cluster. Fig.~\ref{fig:kernelbis} clearly shows that we have no
  specific correlation of WDCOs with bright objects (stars or bright
  galaxies). The density of WDCOs in this area is globally lower than
  in the central regions: the most significant concentrations detected
  in this area show at the 5$\sigma$ level while Fig.~\ref{fig:kernel}
  shows concentrations at the 10$\sigma$ level.

  We therefore believe that our WDCOs are most probably not spurious
  detections and that part of them are likely to be cluster members. 

\begin{figure}[h!]
\begin{center}
\caption{Example of external area around LCDCS~0110. The field shown
  is 1.65' wide and represents a physical area similar to
  Fig.~\ref{fig:kernel} (bottom).  Red circles are the detected
  WDCOs. Blue contours show the WDCO surface density, starting at the
  3$\sigma$ level and are in steps of 1$\sigma$.}
\label{fig:kernelbis}
\end{center}
\end{figure}

Finally, we compared these WDCOs with some very faint galaxies in the
Local Group (e.g.  Massey et al. 2007) to check if these galaxies
have similar stellar populations. The Massey et al. Local Group dwarf galaxies
have typical V-band magnitudes between $-13$ and $-14$, in good agreement
with the dwarf absolute magnitudes computed from our color-magnitude
plots when we assume that WDCOs on the red-sequence are part of
the considered clusters. It is therefore likely that part of these WDCOs are
dwarf galaxies similar to faint Local Group dwarfs.

\section{Summary}

To help us analyse the mechanisms taking place in galaxy clusters, and
place constraints on their formation history and physical properties,
we have searched for intracluster light (ICL) in ten galaxy clusters
at redshifts $0.4<$z$<0.8$. For the first time, we have detected
significant diffuse light sources in an unprecedentedly high redshift
bin z=[0.4,0.8] based on very deep HST ACS images to which we have
applied a very sensitive wavelet detection method.  Until now, most
searches and detailed studies of ICL emission have been limited to
redshifts z$<0.4$ because of the intrinsic faintness of ICL emission,
so our study represents a significant step forward in measuring any
ICL evolution with redshift.

Our analysis has applied a wavelet code (see e.g. Pereira 2003 and
Da Rocha \& Mendes de Oliveira 2005) to deep HST ACS images in the
F814W filter and V-band VLT/FORS2 images (for three of the ten
clusters). Detection levels have been assessed as a function of the
surface brightnesses of the diffuse light sources via simulations.

In the F814W filter, we have detected diffuse light sources in all the
clusters with typical sizes of a few tens of kpc (assuming that the
diffuse light sources are at the cluster redshifts). The ICL detected
by stacking the ten F814W images shows a very clear 8$\sigma$
detection in the source center extending over a $\sim 50\times
50$~kpc$^2$ area. The total absolute magnitude of this source is
$-21.6$ in the F814W filter, equivalent to about two $L^*$ galaxies
for each of the 10 clusters.

We have also discussed the possible anisotropy of the ICL distribution
and the existence of substructures in the inner regions of the
clusters.  We have found a weak correlation between the total F814W
absolute magnitude of the ICL and both the velocity dispersion and the
mass. No correlation was found between the cluster M/L and the amount
of ICL (in agreement with Dolag et al. 2010), and there is no evidence
for any special orientation in the ICL source distributions.  We have
found no strong variation in the amount of ICL between z=0 and
z=0.8, besides perhaps a modest increase.

Finally, besides the extended ICL, we have also found Wavelet-detected
compact objects (WDCOs). Since these sources are very faint, we only
considered those detected in both the HST/ACS/F814W and FORS2/V-band
filters, in the three clusters for which sufficiently deep data in
both bands are available. The fit of a two-dimensional Gaussian plus a
constant background on each of the WDCO images suggests that they are
very unlikely to be faint Galactic stars. On the other hand, part of
the WDCOs are located on the cluster red sequences in color-magnitude
diagrams and their spatial distribution also suggests that they could
be very faint compact galaxies belonging to the considered clusters
and comparable to faint Local Group Dwarfs.

\begin{acknowledgements}
The authors thank the referee for useful and constructive comments.
  We thank the French PNCG/CNRS for support in 2010. We also thank
  A. Cappi, J.G. Cuby, C. Ferrari, J.P. Kneib, R. Malina,
  S. Maurogordato, and C. Schimd for their help.  TS acknowledges
  support from the Netherlands Organization for Scientific Research
  (NWO), from NSF through grant AST-0444059-001, and from the
  Smithsonian Astrophysics Observatory through grant GO0-11147A.

\end{acknowledgements}

\end{document}